\documentclass[twocolumn,trackchanges]{aastex62}

\graphicspath{{./}{figures/}}

\accepted{by ApJ}
\shorttitle{Star formation in NGC 7793}
\shortauthors{Mondal et al.}

\begin{document}

\title{Tracing young star-forming clumps in the nearby flocculent spiral galaxy NGC~7793 with UVIT imaging}

\author{Chayan Mondal}
\affiliation{Indian Institute of Astrophysics, Koramangala II Block, Bangalore-560034, India}
\affiliation{Pondicherry University, R.V. Nagar, Kalapet, 605014, Puducherry, India}

\email{chayan@iiap.res.in, mondalchayan1991@gmail.com}

\author{Annapurni Subramaniam}
\affiliation{Indian Institute of Astrophysics, Koramangala II Block, Bangalore-560034, India}

\author{Koshy George}
%\affiliation{University Observatory Munich, Ludwig-Maximilians-University, Germany}
\affiliation{Faculty of Physics, Ludwig-Maximilians-Universit{\"a}t, Scheinerstr. 1, Munich, 81679, Germany}

\author{Joseph E. Postma}
\affiliation{Dept. Physics and Astronomy, University of Calgary, Calgary, AB, T2N 1N4, Canada}

\author{Smitha Subramanian}
\affiliation{Indian Institute of Astrophysics, Koramangala II Block, Bangalore-560034, India}

\author{Sudhanshu Barway}
\affiliation{Indian Institute of Astrophysics, Koramangala II Block, Bangalore-560034, India}

\keywords{galaxies: spiral – galaxies: individual – galaxies: star formation}
\begin{abstract}
Star formation in galaxies is a hierarchical process with a wide range of scales from smaller clusters to larger stellar complexes. Here, we present an ultra-violet imaging study of the nearby flocculent spiral galaxy NGC 7793, observed with the Ultra-Violet Imaging Telescope (UVIT). We find that the disk scale-length estimated in Far-UV (2.64$\pm$0.16 kpc) is larger than that in Near-UV (2.21$\pm$0.21 kpc) and optical (1.08 kpc), which supports the inside-out growth scenario of the galaxy disk. The star-forming UV disk is also found to be contained within the extent of H~I gas of column density greater than $10^{21}$cm$^{-2}$. With the spatial resolution of UVIT (1 pixel $\sim$ 6.8 pc), we identified 2046 young star-forming clumps in the galaxy with radii between $\sim$ 12 - 70 pc, which matches well with the size of GMCs detected in the galaxy. Around 61\% of the regions identified in our study have age younger than 20 Myr, which points to a recent enhancement of star formation across the galaxy. We also noticed that the youngest star-forming regions, with age $<$ 10 Myr, distinctly trace the flocculent arms of the galaxy. The estimated mass of the clumps cover a range between $10^3 - 10 ^6 M_{\odot}$. We noticed a gradient in the mass distribution of identified clumps along the spiral arms. We have also studied the nuclear star cluster of the galaxy and found that the stellar populations in the cluster outskirts are younger than the inner part. 
\end{abstract}

\section{Introduction}

Galaxies of different morphology have distinct nature of ongoing star formation. The massive ellipticals, systems that have exhausted its molecular gas, show no signs of star formation, whereas gas-rich spiral and irregular galaxies are found to be more active in nature. Star formation in spiral galaxies is observed mainly in their disks containing the spiral arms \citep{blanton2009}. Depending on the nature of arms, \citet{elmegreen1981} classified spirals into two categories. Galaxies having long and continuous well-defined arms with two-armed axial symmetry are called grand design spirals, whereas there are flocculent spirals with patchy, fragmented, short wispy spiral structures. Understanding the formation and evolution of these different spiral features is important in the context of galaxy evolution. 

The observed morphology of spiral features found to show variations in different wavebands as stars formed in the density waves start moving out gradually with increasing age \citep{hamed2016}. The role of density waves in triggering star formation along the spiral arms is studied with both simulations and observations. \citet{roberts1969} proposed that spiral density waves can induce gravitational collapse of the gas clouds and trigger star formation along the arms. Although the efficiency of star formation locally depends on several factors such as gas density, gravitational potential, shear as well as coriolis forces due to disk rotation, ambient pressure, and metallicity of the ISM \citep{leroy2008}. \citet{elmegreen1986} reported only a small difference between the average star formation rate (SFR) in grand design and flocculent spiral galaxies. This indirectly says that spiral density wave does not have a strong impact to trigger star formation in galaxies. Several  observational studies (for example, \citet{seigar2002}) have supported the idea of \citet{roberts1969} for triggered star formation in spirals. Observation in FUV or 8 $\mu$m, which traces the location of youngest stars or the ongoing star formation in a galaxy, delineates the current morphology of spiral arms, whereas optical and near-infrared bands pick up the relatively older stars which have moved out from the density waves \citep{hamed2016}. 

Spiral galaxies mostly follow inside-out growth scenario, where the inner part of the disk form earlier than the outer part \citep{white1991,mo1998,brook2006}. Several studies, that examined the star formation history across the disk of many nearby spirals, have found evidence of inside-out formation \citep{mateos2007,gogarten2010}. \citet{mateos2007} studied 161 nearby spiral galaxies with GALEX and 2MASS observations and found signature of moderate inside-out disk growth for the majority of the sample. Several studies also reported the radial migration of stars as one of the reasons for the observed flattening of metallicity or age of stellar populations in the outer disk of spiral galaxies \citep{roskar2008,vlajic2009,vlajic2011}.

NGC~7793 is a SA(s)d type flocculent spiral galaxy of the nearby sculptor group \citep{devaucouleurs1991}. It is located at a distance of 3.4 Mpc and has two nearby dwarf companions \citep{zgirski2017,koribalski2018}. The galaxy has an absolute B band magnitude of $-18.31$ and stellar mass of 3.2$\times$10$^9$ M$_{\odot}$ with a sub-solar metallicity \citep{carignan1985,bothwell2009,vandyk2012}. The optical radius ($R_{25}$) of the galaxy is around 4.67$^{\prime}$ ($\sim$4.62 kpc) \citep{devaucouleurs1991}. The properties of the galaxy are listed in Table \ref{ngc7793}. \citet{elmegreen1984} identified NGC~7793 as an extreme flocculent galaxy and reported that the galaxy does not show specific structures in its older disk, whereas images in the bluer band show structures which are possibly due to star formation or weak stellar density ripples. The galaxy is reported to have a very small bulge and a nuclear star cluster at the center \citep{dicaire2008,kacharov2018}. 

Several observational studies have explored the disk properties of NGC~7793 in various wavebands. \citet{carignan1985} studied the H~I disk of the galaxy and reported it to be extended up to 1.5 times the optical diameter of the galaxy. They also noticed the galaxy rotation curve to be declining in nature in the outer part. A recent H~I study by \citet{koribalski2018} traced the H~I disk up to further out, along with a significant wrap in the outer disk. The properties of the CO molecular gas of the galaxy is also studied by \citet{muraoka2016}. \citet{hermanowicz2013} used GALEX FUV data to identify several star-forming regions in the galaxy and compared their FUV flux to that with H$\alpha$. \citet{dicaire2008} identified the signature of H$\alpha$ emission up to the extent of the H~I disk, and this points to the ongoing star formation across the galaxy. Several studies are also done with HST observations to understand the star formation history, hierarchy of star-forming regions, and the interplay between young star clusters and giant molecular clouds in this galaxy \citep{elmegreen2014,radburn2012,grasha2018,sacchi2019}. These observations cover a wide wavelength range from near-UV to infrared. The results conclude that the disk of the galaxy supports an inside-out growth scenario with a considerable amount of stars formed in the recent time. \citet{radburn2012} reported that the outer disk of the galaxy beyond $\sim$ 3 kpc is mostly populated with younger populations, which show a break at radius $\sim$ 5 kpc.

The young stellar populations of a galaxy emit most of their radiation in the ultra-violet (UV) band \citep{kennicutt2012}. Hence, imaging a galaxy in FUV and Near-UV (NUV) will help to locate the regions with active star formation and to delineate the overall disk structure as traced by the younger population. In this study, we explored the nature of UV disk and the properties of young star-forming regions of the galaxy NGC~7793 with FUV and NUV broadband observations from the Ultra-Violet Imaging Telescope (UVIT). The primary aim is to explore the recent star-forming activity in this flocculent spiral system, which offers a unique platform to investigate the characteristics of density waves. With the large field of view of UVIT, we are able to cover the entire galaxy disk well beyond its optical radius. This provides us the scope to understand the overall disk structure in UV and also to pick up young star-forming clumps across the disk including the outskirts. Though the disk of the galaxy has been explored in different wavebands, a detailed study has not been done by combining the data from both FUV and NUV band together. Our study provides a comprehensive view of the young star-forming regions in the galaxy resolved up to length scales of $\sim$ 25 pc. We have analysed the overall UV disk profile and correlated it with the H~I column density map of the galaxy. The deep and high-resolution UV images are utilised to identify star-forming clumps and estimate their age and mass using simple stellar population (SSP) models. The study brings out the spatial distribution of young stellar clumps as a function of their age and mass across the entire disk. We have also analysed the nuclear star cluster of the galaxy using the UVIT data. The paper is arranged as follows. The observations and data are presented in Section \ref{s_observation}, theoretical models in Section \ref{s_model}, extinction in UV in Section \ref{s_extinction} and analysis in Section \ref{s_analysis} followed by results and discussions and a summary in Sections \ref{s_results} and \ref{s_summary} respectively.

\begin{table}
\centering
\caption{Properties of NGC~7793}
 \label{ngc7793}
\resizebox{90mm}{!}{
\begin{tabular}{ccc}
\hline
 Property & Value & Reference\\\hline
 Morphological type & SA(s)d & \citet{devaucouleurs1991}\\
 RA & 23 57 49.7 & \citet{skrutskie2006}\\
 DEC & -32 35 27.6 & \citet{skrutskie2006}\\
 Distance & 3.4 Mpc & \citet{zgirski2017}\\
 Metallicity (Z) & $0.6Z_{\odot}$ & \citet{vandyk2012}\\
 Inclination & $53.7^\circ$ & \cite{carignan1985}\\
 PA of major axis & $279.3^\circ$ & \citet{carignan1985}\\
 Stellar mass & 3.2$\times$10$^9$ M$_{\odot}$ & \citet{bothwell2009}\\\hline
\end{tabular}
}
\end{table}

\begin{figure*}
    \centering
    \includegraphics[width=6.0in]{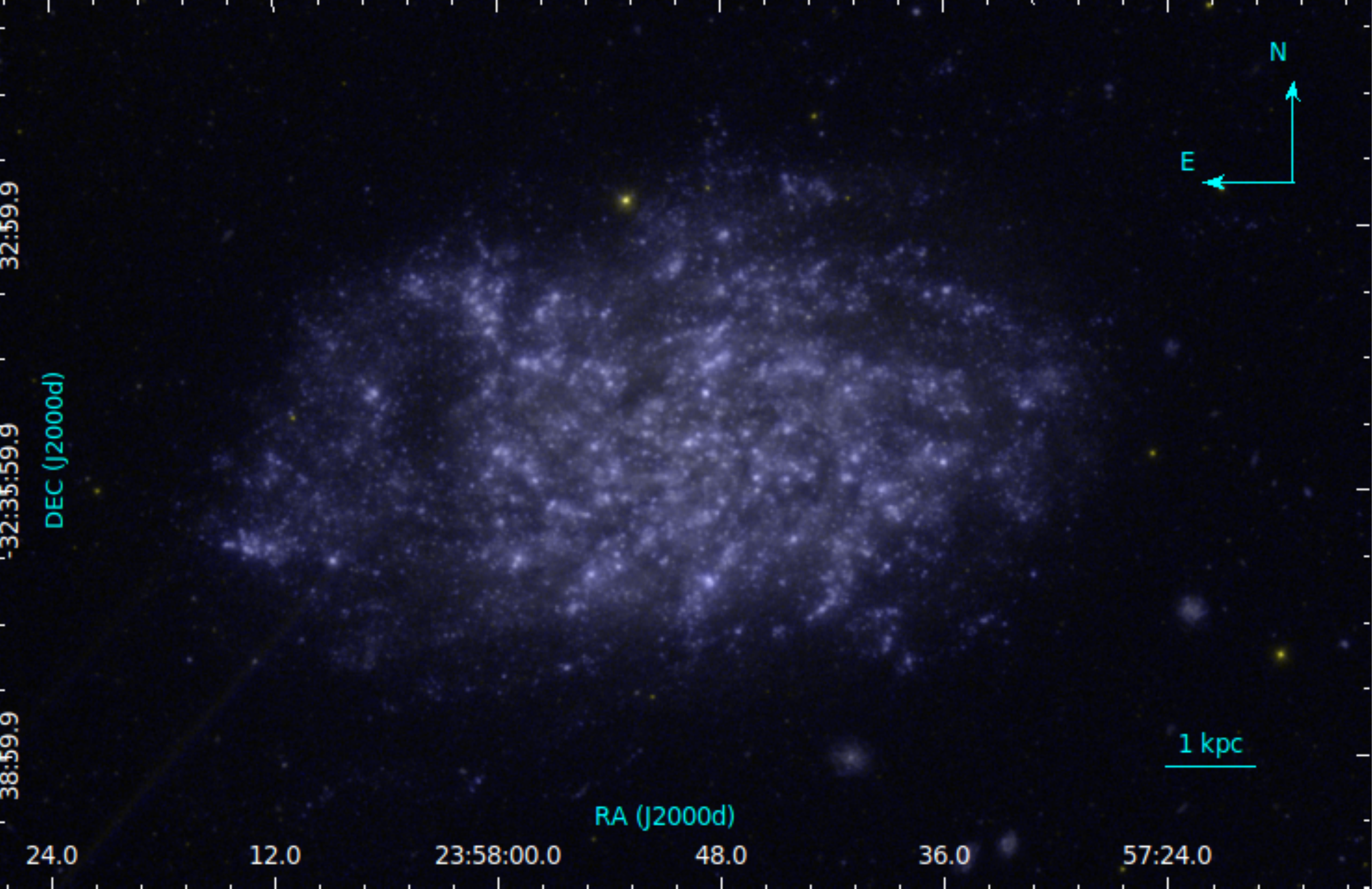}
    \caption{The UVIT color composite image of the galaxy NGC~7793. The emission in F148W and N242W filter are represented by blue and yellow color respectively. }
    \label{ngc7793_color}
\end{figure*}

\section{Observations and Data} 
\label{s_observation}
The galaxy NGC~7793 was imaged in UV with Ultra-Violet Imaging Telescope (UVIT) onboard AstroSat satellite \citep{kumar2012} (Figure \ref{ngc7793_color}). UVIT has the capability to observe simultaneously in FUV, NUV, and visible bands. Each of the FUV and NUV channels of the telescope is equipped with multiple filters. The visible channel is used to track the drift of the satellite during the course of observations. Along with the multi-band imaging in UV, the telescope also offers a superior spatial resolution of $\sim1.4^{\prime\prime}$, which altogether makes the instrument unique for imaging in UV. NGC~7793 was observed in two UVIT broad band filters F148W (FUV) and N242W (NUV). The complete observation was performed with 8 orbits of the satellite on 10 November, 2016. We used CCDLAB software \citep{postma2017} to correct the drift for all the images. Each image is flat-fielded and further corrected for distortion, and fixed pattern noise using the calibration files \citep{girish2017,postma2011}. The corrected images are co-aligned and combined with the help of same software to produce final deep images. The final images have 4096$\times$4096 pixel dimension with 1 pixel corresponds to $\sim$0.4$^{\prime\prime}$ ($\sim$ 6.8 pc at the distance of NGC~7793). The exposure times obtained for the images in F148W and N242W band are $\sim$ 6.5ks and 9ks respectively. The zero point magnitude and the unit conversion factors (Table \ref{uvit_obs}) for the used filters are adopted from \citet{tandon2017}. We also used H~I column density map of the galaxy from The H$~$I Nearby Galaxy Survey (THINGS) \citep{walter2008}. 

\begin{table*}
\centering
\caption{Details of UVIT observations}
\label{uvit_obs}
\begin{tabular}{p{2cm}p{2cm}p{3cm}p{3.5cm}p{2cm}p{3cm}}
\hline
Filter & Bandpass & ZP magnitude & Unit conversion & $\triangle\lambda$ & Exposure time\\
 & ($\AA$) & (AB) & (erg sec$^{-1}$cm$^{-2}\AA^{-1}$) & ($\AA$) & (sec)\\\hline
F148W & 1250-1750 & 18.016 & 3.09$\times10^{-15}$ & 500 & 6543\\
N242W & 2000-3000 & 19.81 & 2.22$\times10^{-16}$ & 785 & 8919\\\hline
\end{tabular}
\end{table*}

\begin{table}
\centering
 
\caption{Starburst99 model parameters}
\label{starburst99_t}
\resizebox{90mm}{!}{
\begin{tabular}{cc}
\hline

 Parameter & Value\\\hline
 Star formation & Instantaneous\\
 Stellar IMF & Kroupa (1.3, 2.3)\\
 Stellar mass limit & 0.1, 0.5, 120 $M_{\odot}$\\
 Total cluster mass & $10^3 M_{\odot}$-$10^6 M_{\odot}$\\
 Stellar evolution track & Geneva (high mass loss)\\
 Metallicity & Z=0.008\\
 Age range & 1-900 Myr\\ \hline

\end{tabular}
}

\end{table}

\section{Theoretical models}
\label{s_model}
\begin{figure}
    \centering
    \includegraphics[width=3.7in]{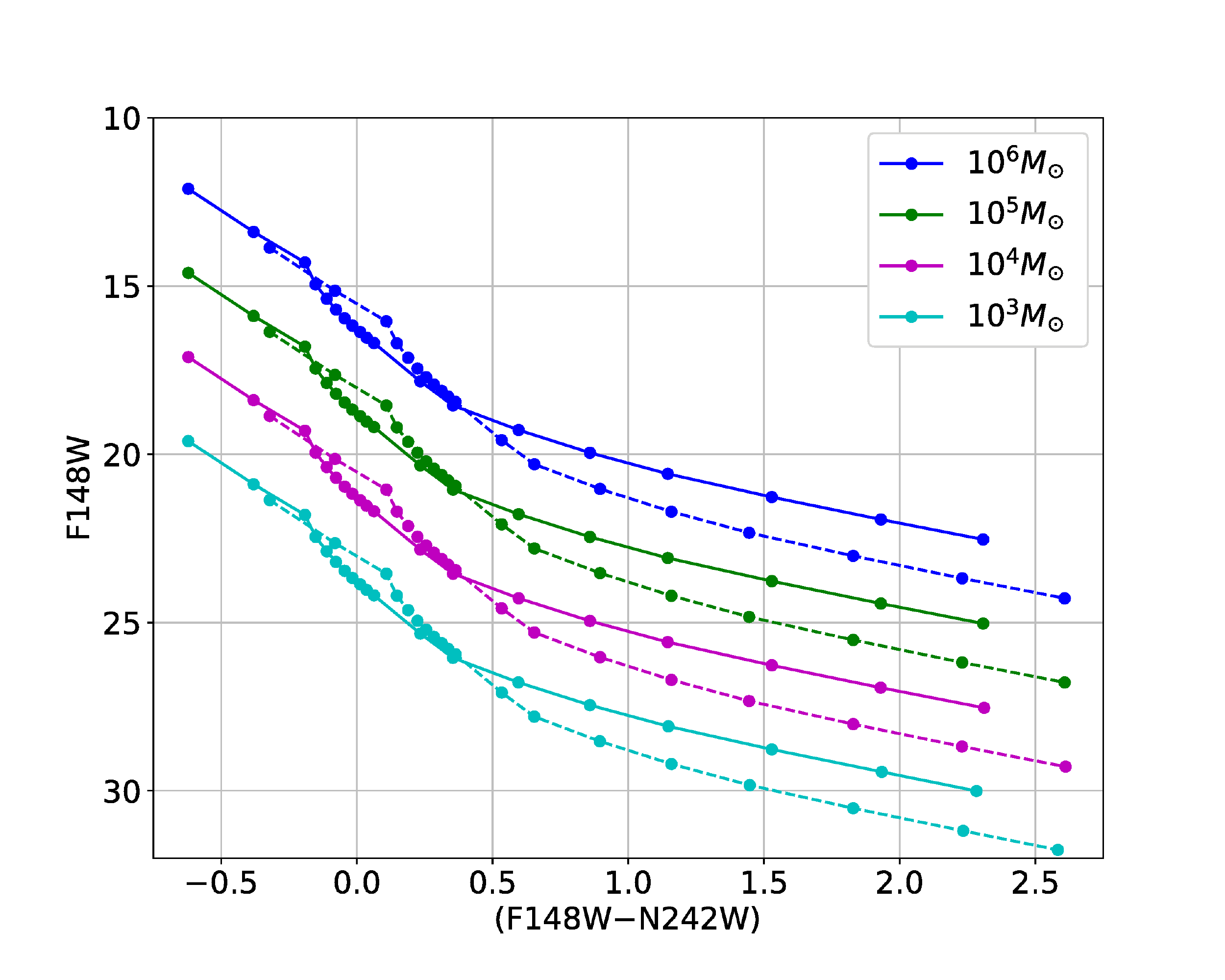}
    \caption{F148W vs (F148W$-$N242W) color-magnitude diagram, simulated using Starburst99 SSP model. Different curves (continuous line) signify four different
    total cluster masses ($10^6 M_\odot$, $10^5 M_\odot$, $10^4 M_\odot$, $10^3 M_\odot$). The points shown in each curve are for different ages starting from 1 Myr to 900 Myr 
(increasing along the color axis) with age interval 10 Myr for 1 to 100 Myr range and 100 Myr for 100 to 900 Myr. The dashed lines are plotted by adding the extinction and reddening, as applicable in the galaxy NGC~7793 (Section \ref{s_extinction}), to the model curves. The values of model input parameters are listed in Table \ref{starburst99_t}.}
    \label{starburst99}
\end{figure}

In order to characterize the star-forming regions in NGC~7793, we used the Starburst99 SSP model \citep{leitherer1999}. The model offers integrated spectra of young star clusters for a set of chosen parameters. These spectra can be used to produce diagnostic diagrams for tracing the evolutionary stages of an unresolved star cluster in galaxies. We exploit Starburst99 model data and produced Figure \ref{starburst99} to estimate the age and mass of star-forming clumps in the galaxy. We assumed instantaneous star formation law and kroupa stellar initial mass function \citep{kroupa2001} for the stellar-mass limit of 0.1 - 120 $M_{\odot}$. We considered 19 spectra between the age range 1 - 900 Myr and metallicity Z = 0.008. The metallicity is adopted to be the closest to that of NGC~7793, as reported by \citet{vandyk2012} among the available model values. We convolved UVIT filter effective area with these spectra and estimated the expected magnitudes in F148W and N242W bands for four different cluster masses ($10^6 M_\odot$, $10^5 M_\odot$, $10^4 M_\odot$, $10^3 M_\odot$) at the distance of the galaxy. The spectra we considered in our study include flux from both stellar and nebular emission. The simulated plot (Figure \ref{starburst99}) signifies that (F148W$-$N242W) color of a cluster changes only with its age, whereas for a fixed age F148W magnitude becomes brighter with increasing mass and vice versa. Therefore, the magnitude axis of Figure \ref{starburst99} will trace the mass of a cluster, and the age can be estimated from the observed color value.

\section{Extinction in UV}
\label{s_extinction}
One of the important factors that need to be considered while studying an external galaxy in UV is the effect of extinction. This is because of the higher value of the extinction coefficient in UV and also due to the characteristic variation of extinction law in different galaxies. In order to estimate the extinction value in UVIT filters, we adopted the value of $E(B-V) = 0.179$ for NGC~7793 from the study of \citet{bibby2010}. Also, as the metallicity of the galaxy is similar to the Large Magellanic Cloud (LMC), we assumed the extinction law to be average LMC type as modeled by \citet{gordon2003}. We used the extinction law calculator of \citet{mccall2004} supplied in NASA/IPAC Extra-galactic Database (NED) to estimate the value of extinction coefficients ($R_{\lambda}$) in F148W and N242W bands. The values of $R_{F148W}$ and $R_{N242W}$ are found to be 9.78 and 8.12 respectively. Using the value of $R_{\lambda}$ and $E(B-V)$ in equation \ref{eq_ext_7793}, we estimated the value of extinctions in both the bands. The values of $A_{F148W}$ and $A_{N242W}$ are found to be 1.75 mag and 1.45 mag respectively. We used these values to correct the observed flux of the identified objects in our study. 

\begin{equation}
A_{\lambda} = R_{\lambda}E(B-V)  
\label{eq_ext_7793}
 \end{equation}
 
\section{Analysis} 
\label{s_analysis}
\subsection{UV disk profile of NGC~7793}
\label{uv_disk_s}
\begin{figure}
    \centering
    \includegraphics[width=3.7in]{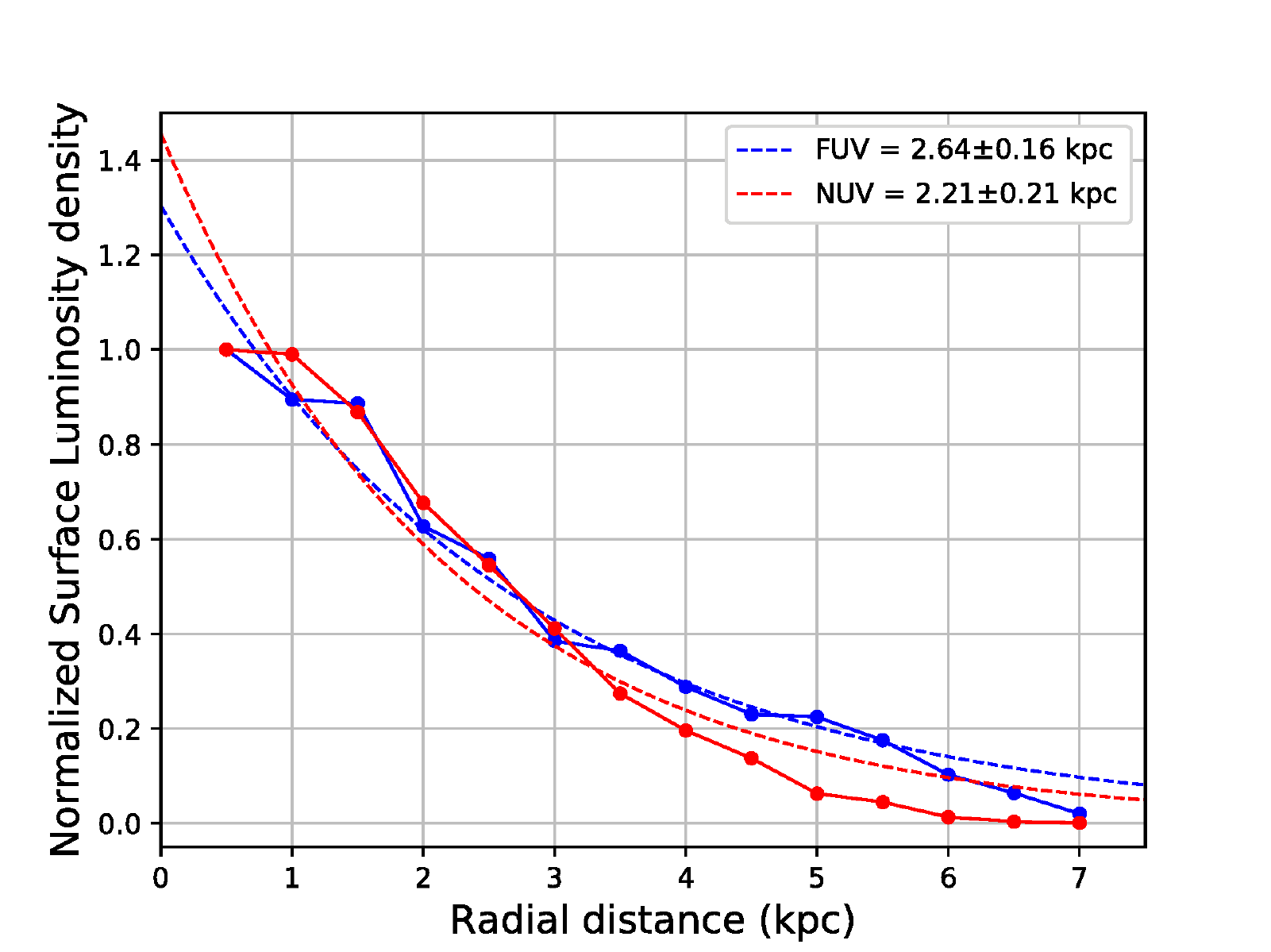}
    \caption{The FUV and NUV radial luminosity density profiles of the galaxy are respectively shown in solid blue and red lines. Each profile is normalized with respect to the maximum of the respective curve. The dashed lines are the exponential fits for each observed profiles. The values of the estimated disk scale-length in FUV and NUV are 2.64$\pm$0.16 kpc and 2.21$\pm$0.21 kpc respectively.}
    \label{luminosity}
\end{figure}
The FUV and NUV emission in a galaxy primarily trace the stellar population of age up to 100 and 200 Myr respectively \citep{kennicutt2012}. In order to understand the characteristics of UV emission in the galaxy, we produced radial luminosity density profiles in both FUV and NUV. The nature of these profiles is expected to highlight the distribution of younger star-forming regions across the galaxy. We assumed the distance, galaxy center, inclination, and position angle of the galaxy from Table \ref{ngc7793} and used the equation given in section 2 of \citet{marel2001} to estimate galactocentric distance in kpc for each image pixel. Starting from the galaxy center, we considered concentric annuli of width 0.5 kpc and estimated the value of luminosity density (erg sec$^{-1}$pc$^{-2}$) in each annulus from the measured flux. We normalized the measured luminosity values with respect to the maximum of the respective curve and plotted the profiles for both FUV and NUV in Figure \ref{luminosity}. Both the profiles show exponential nature. 

In order to estimate the disk parameters, we fitted exponential curves to these observed profiles and estimated the values of disk-scale length ($R_{d}$) in FUV and NUV. The values of $R_{d}$ are found to be 2.64$\pm$0.16 kpc and 2.21$\pm$0.21 kpc respectively, in FUV and NUV. The optical disk scale-length of the galaxy is reported to be 1.08 kpc by \citet{carignan1985}. Hence the disk of the galaxy is more extended towards shorter wavelengths. We noticed both the observed profiles in Figure \ref{luminosity} to follow each other up to radius 3 kpc. Beyond that, the profile of FUV luminosity becomes flatter than the NUV in the outer disk. These together signify that the stellar populations detected in UV in the outer disk beyond 3 kpc are mostly younger. Our results satisfy the inferences projected in \citet{radburn2012}. Using HST observations, \citet{radburn2012} identified stars of different ages and found that the radial number density profile of younger stars between 3 - 5 kpc is flatter compared to that of the older one. We further noticed that the observed FUV luminosity density profile has a break at a radius of around 5 kpc, which is also noticed in Figure 6 of \citet{radburn2012} for the younger populations.

\subsection{Correlation with H~I column density}
\begin{figure*}
    \centering
    \includegraphics[width=6in]{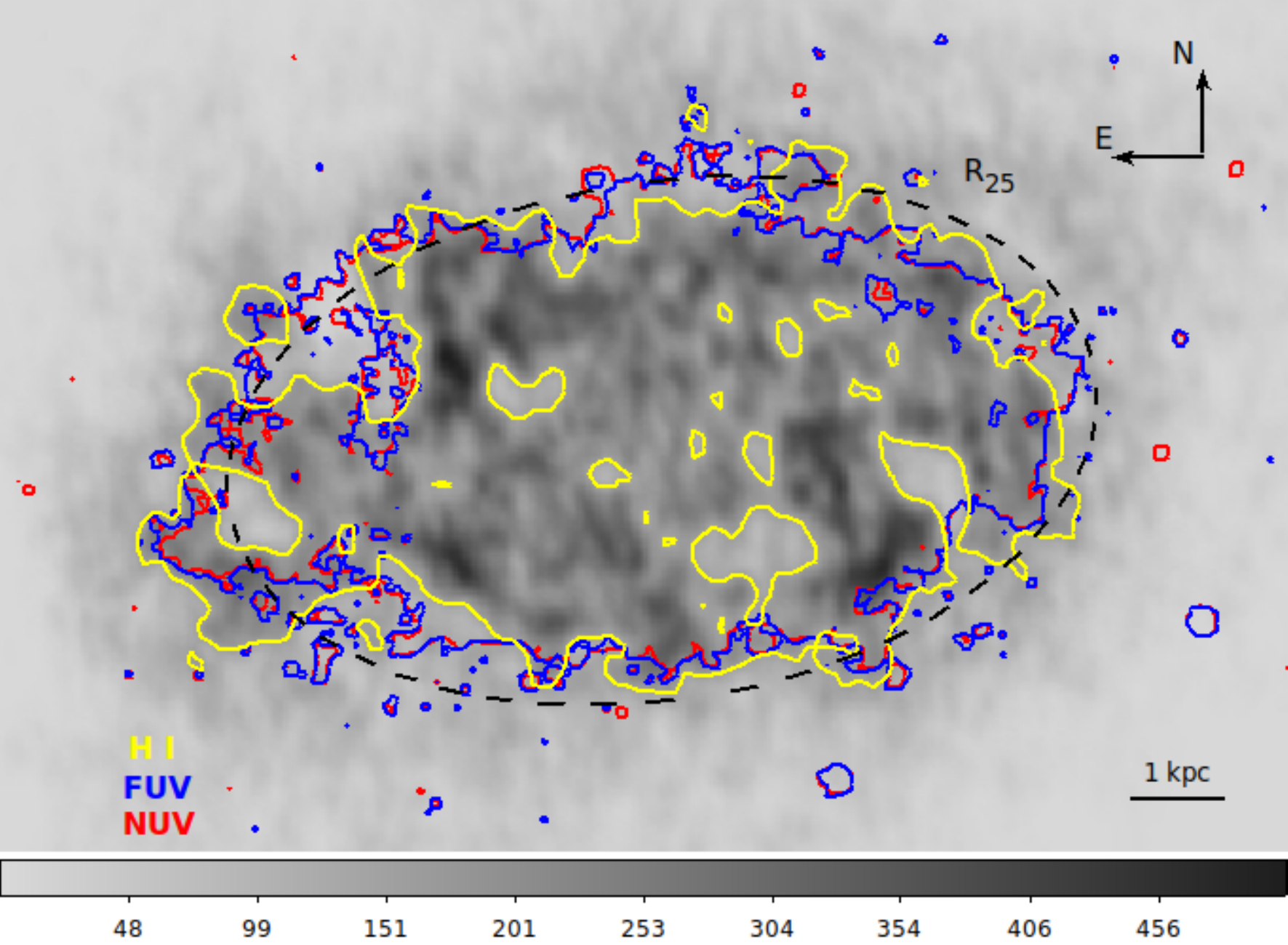}
    \caption{The figure shows the H~I moment 0 map of the galaxy. The grey scale is in Jy/B*M/S. The yellow contours show regions with H~I column density greater than $10^{21} cm^{-2}$. The blue and red contours respectively represent FUV and NUV emission profiles. These are generated for a threshold flux of 5 times the average background value measured in each respective band. The black dashed ellipse shows the $R_{25}$ boundary of the galaxy.}
    \label{hi}
\end{figure*}
The H~I column density is known to have a threshold value of $\sim$ $10^{21} cm^{-2}$ for star formation in galaxies \citep{skillman1987,clark2014}. \citet{carignan1990} observed NGC~7793 with VLA and traced H~I gas of column density $5\times10^{19} cm^{-2}$ up to 1.5$R_{25}$ radius of the galaxy. As the FUV and NUV disk emission profiles trace the distribution of young star-forming regions, we compare those with H~I gas density profile to understand the relation between gas and star formation. In Figure \ref{hi}, we have shown the moment 0 H~I map of the galaxy, observed with VLA by \citet{walter2008}. To compare the emission profiles in FUV, NUV, and H~I, we plotted contours as displayed in Figure \ref{hi}. The yellow contour signifies the extent of the H~I disk with column density more than $10^{21} cm^{-2}$. To trace the extent of the disk in FUV and NUV, we adopted the threshold as 5 times the average background flux in each respective band and created contours. The contours are shown in blue and red respectively for FUV and NUV in the same figure. The extent of the disk emission in both FUV and NUV is found to be almost similar for the adopted thresholds. Also, the H~I disk with column density more than $10^{21} cm^{-2}$ closely matches with both the UV profiles. In some regions along the east and western part, we noticed H~I contours to be a little more extended than the UV. In the northern part, we found a part of UV contours to extend outside the H~I contour. The overall good spatial correlation between UV and H~I profiles signifies that star formation is happening in the galaxy up to the extent where H~I column density is more than the threshold value $10^{21} cm^{-2}$.

\begin{figure}
    \centering
    \includegraphics[width=3.7in]{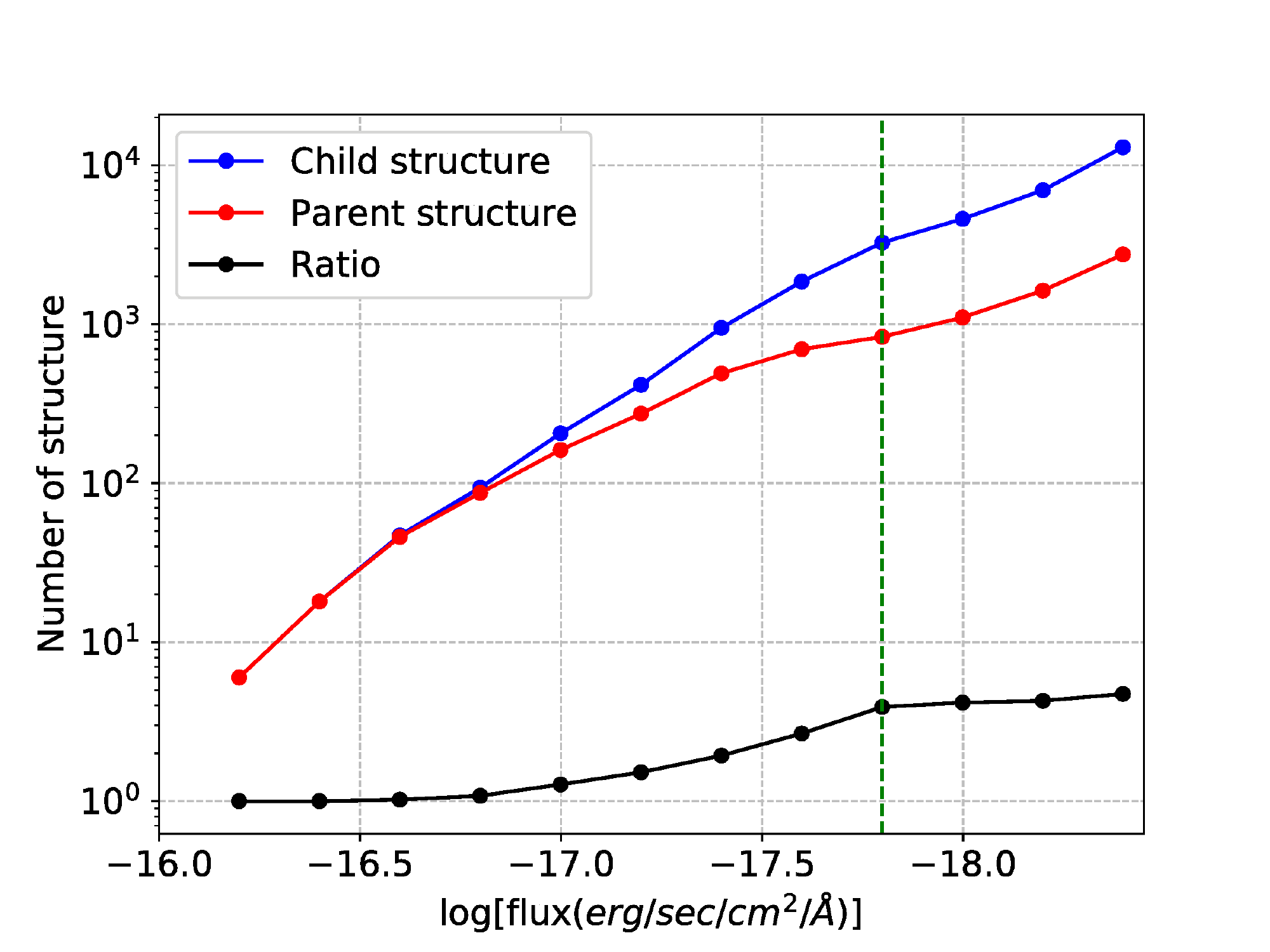}
    \caption{The number of identified parent and child structures are shown as a function of varying threshold flux. The black line shows the ratio of the number of child and parent structures. The vertical green dashed line represents the threshold flux (log(flux) = $-17.80$) selected for our analysis.}
    \label{fuv_lt}
\end{figure}

\begin{figure}
    \centering
    \includegraphics[width=3.4in]{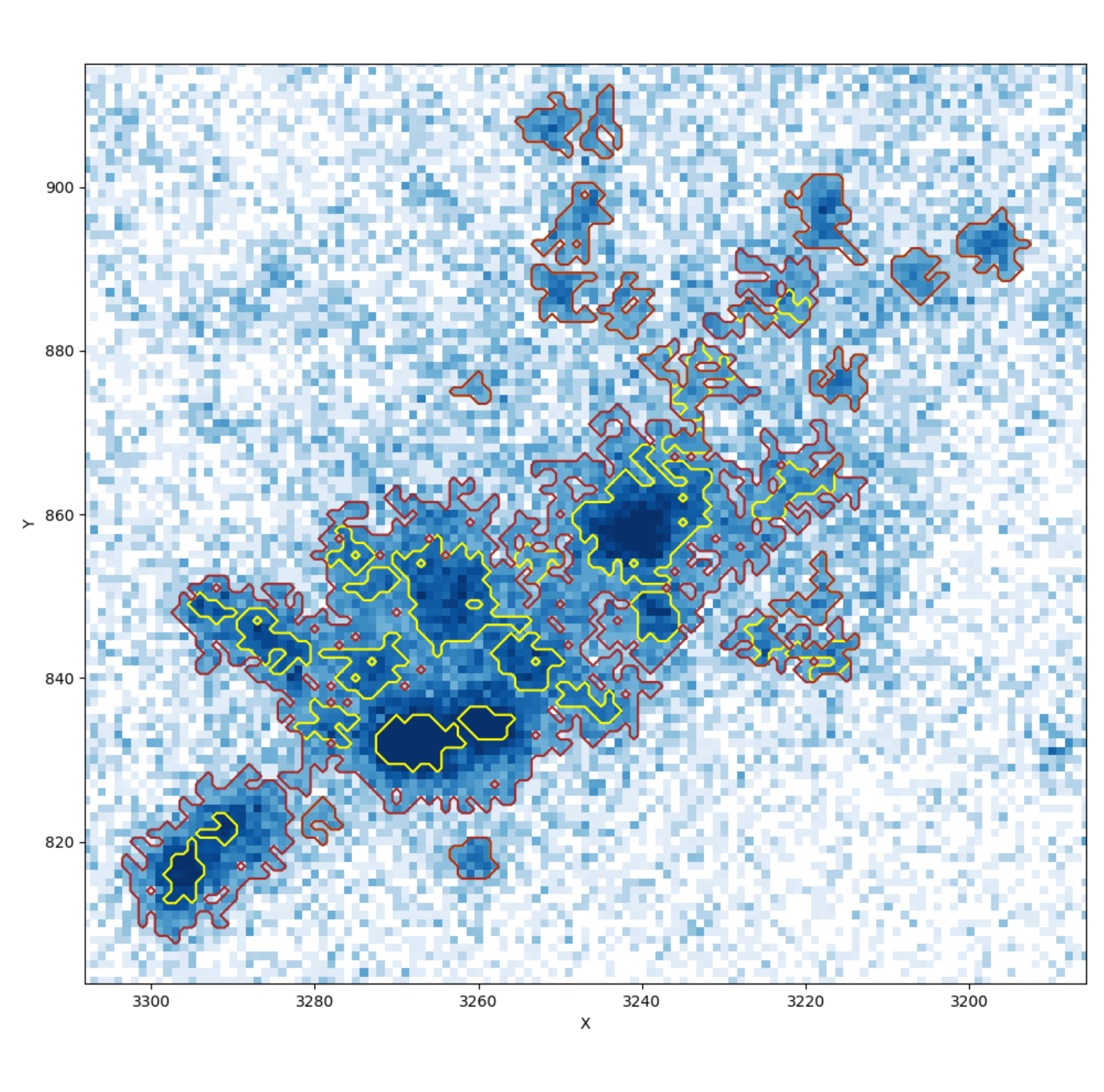}
    \caption{A particular star-forming region of the galaxy is shown along with the contour of parent (brown) and child (yellow) structures. The background shows UVIT F148W band image. The value of the adopted threshold flux is log[flux(erg sec$^{-1}$cm$^{-2}\AA^{-1}$)] = $-17.80$.}
    \label{structure}
\end{figure}

\subsection{Identification of young star-forming clumps}
\label{clumps_7793_s}
As the FUV emission is predominantly contributed by massive young OB stars, imaging in FUV will eventually trace the young star-forming regions hosting massive stars in a galaxy. To identify young star-forming clumps in NGC~7793, we utilized the UVIT FUV observation. We used \textit{astrodendro}\footnote{http://www.dendrograms.org/} python package to locate bright star-forming clumps in the FUV image. The code helps to identify clumps in an intensity map for a given value of threshold flux and a minimum number of pixel. It locates both parent and child structures and builds the dendrogram. We fixed the minimum number of pixel value as 10, which means structures formed with less than 10 pixels are not considered. The area covered by 10 pixels is equivalent to a circle of radius $\sim$ 1.8 pixel ($\sim$ 12 pc at the distance of the galaxy). This value is chosen such that the size of the smallest detectable clump ($\sim$3.6 pixel) matches with the FWHM of the PSF, which is $\sim$ 3.5 pixel in the observed field. 

To fix the threshold flux, we started with a value 5 times the average FUV background flux (log[flux(erg sec$^{-1}$cm$^{-2}\AA^{-1}$)] = $-18.40$) and increased it with 0.2 interval in logarithmic scale up to log(flux) = $-16.20$. For each threshold, we counted the number of identified parent and child structures and plotted in Figure \ref{fuv_lt}. The individual structures with no sub-structure inside are counted as both parent and child structure in \textit{astrodendro}. The ratio of the child and parent structures, represented by the black line, shows an increasing trend with decreasing threshold flux and becomes nearly constant after the value log(flux) = $-17.80$. This flux value also corresponds to the flux of a B5 spectral type star at the distance of NGC~7793. As stars with spectral type cooler than B type does not contribute much of FUV flux, we fixed this value (log[flux(erg sec$^{-1}$cm$^{-2}\AA^{-1}$)] = $-17.80$) to be the threshold flux. For this threshold, the number of identified parent and child structures are found to be 835 and 3266 respectively, within the galactocentric radius of 7 kpc. We considered 7 kpc as the outer boundary, as this value is equivalent to $\sim$ 1.5$R_{25}$, up to which \citet{carignan1990} traced the H~I disk. Again, as per the study of \citet{radburn2012}, the signature of the younger population has been traced up to this radius with HST observations. In Figure \ref{structure}, we have shown an example of parent and child structures identified at a particular star-forming region of the galaxy for the chosen threshold flux. 

\begin{figure}
    \centering
    \includegraphics[width=3.6in]{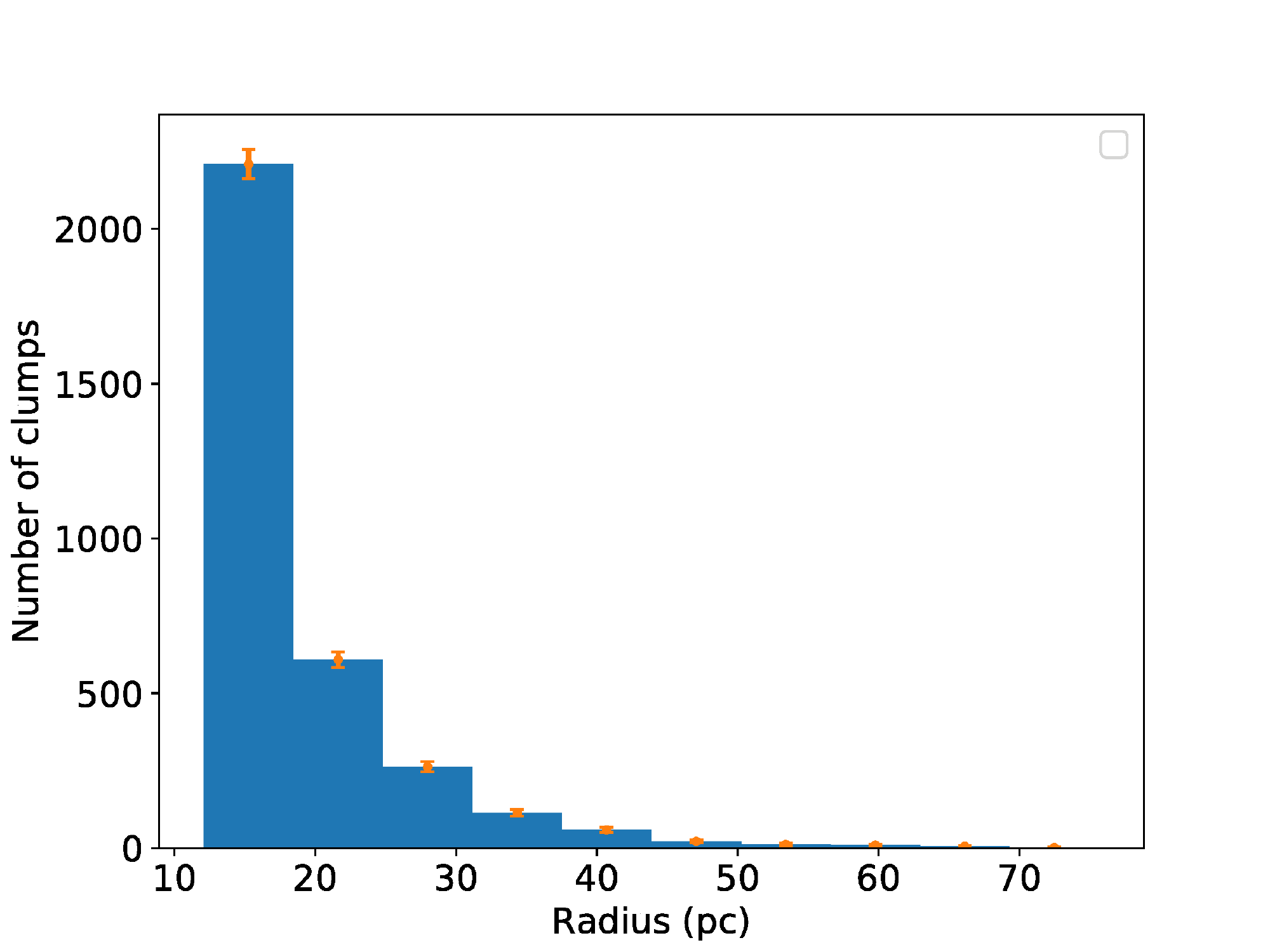}
    \caption{The histogram shows the size of identified clumps (child structure).}
    \label{fuv_leaves_hist}
\end{figure}

As the larger parent structures contain multiple unresolved clumps inside, we considered the child structures, which are identified as individual star-forming clumps in the UVIT image, for our further analysis. The output of \textit{astrodendro} provides position, area, and flux of each identified clumps. We considered the area of these irregular shaped clumps and equated it to the area of a circle to estimate the equivalent radius, which shows a range between $\sim$ 12 - 70 pc. The histogram of the size is shown in Figure \ref{fuv_leaves_hist}. The error bars shown in the histogram are measured from the square root of the numbers in each bin. This matches well with the size of the giant molecular clouds (GMCs) detected in the galaxy \citep{grasha2018}. We used the position and area equivalent radius of these identified clumps as measured in the FUV image and estimated FUV and NUV fluxes from each respective images for the same calculated aperture size. This provides both FUV and NUV magnitudes of the clumps. We corrected these magnitudes for background and extinction. The average background in each individual image is estimated by considering four circular regions, each of radius 1 arcmin, in the observed field located away from the galaxy. The extinction correction is done as explained in section \ref{s_extinction}. Several studies also explored the maximum size of the star-forming region over which star formation is physically correlated \citep{grasha2017}. \citet{sanchez2010,grasha2017,rodriguez2020} used the detected stars and clusters in selected nearby galaxies and reported a length scale between 200 to 1000 pc for such largest hierarchical structures. In this study, we used the FUV flux map to identify individual clumps or large structures in the observed image.  With the adopted threshold flux (i.e., log[flux(erg sec$^{-1}$cm$^{-2}\AA^{-1}$)] = $-17.80$), we found the largest structure formed by the star-forming clumps has size $\sim$ 3 kpc in the galaxy. The size is estimated from the area of the largest parent structure as identified by {\it astrodendro}.

\begin{figure}
    \centering
    \includegraphics[width=3.6in]{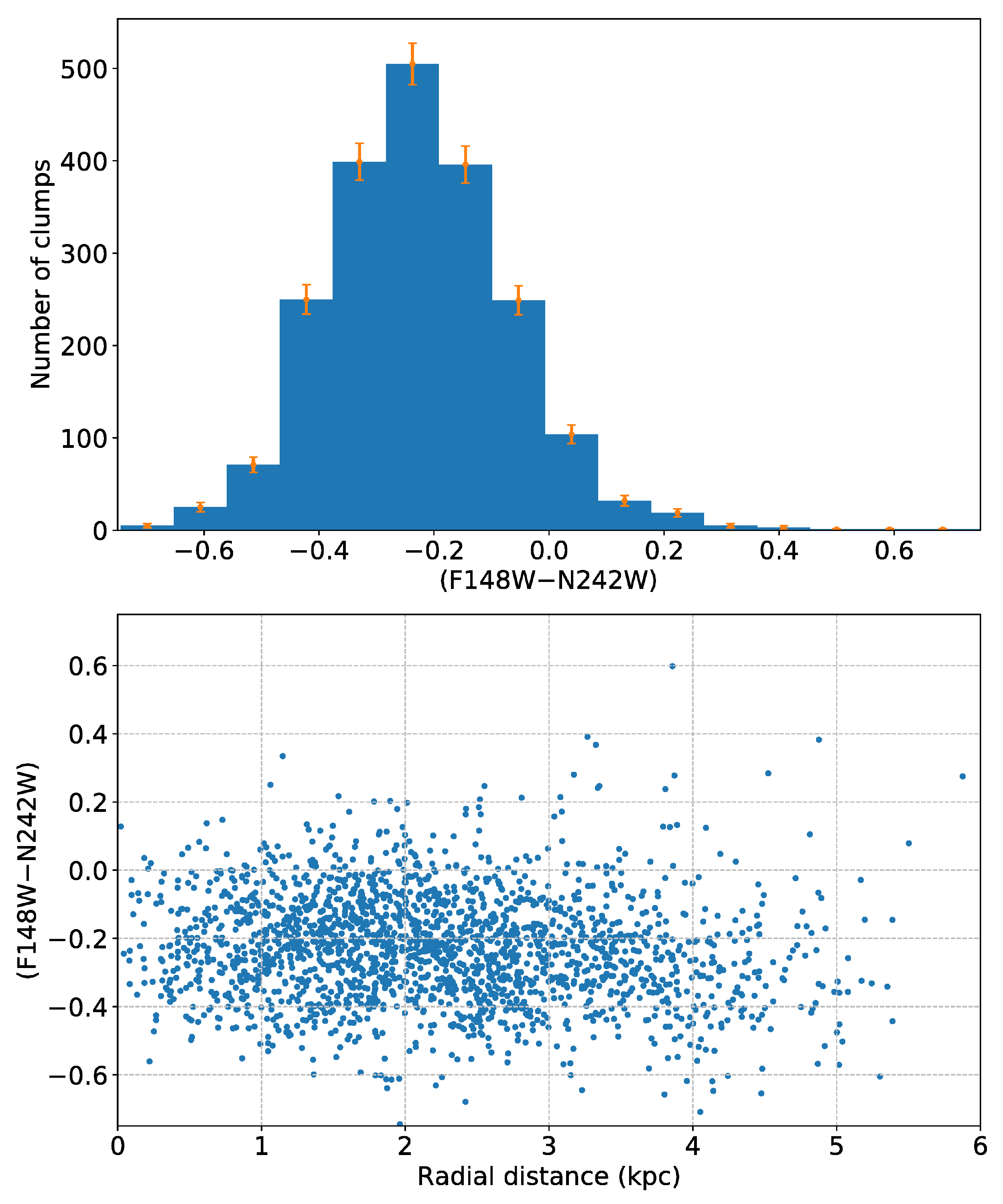}
    \caption{The histogram of (F148W$-$N242W) color of the clumps brighter than 21 mag in F148W band is shown in the upper panel. In the lower panel, we have shown their distribution as a function of radial distance from the galaxy centre.}
    \label{color_hist}
\end{figure}

In order to avoid clumps with relatively larger photometric error ($\gtrsim$0.1 mag), we considered only those clumps with corrected FUV magnitude brighter than 21 mag in our analysis. This results in 2046 number of individual star-forming clumps. We showed the histogram of (F148W$-$N242W) color of these clumps in the top panel of Figure \ref{color_hist}. The distribution shows a gaussian nature with a peak around (F148W$-$N242W) = $-0.25$. We also estimated the galactocentric distance for each clump by adopting the method, as explained in Section \ref{uv_disk_s}. The (F148W$-$N242W) colors of the clumps are plotted as a function of galactocentric distance in the bottom panel of Figure \ref{color_hist}. The distribution shows a flat nature for the inner 3 kpc radius of the galaxy, whereas the clumps identified in the outer part beyond 3 kpc display a slightly bluer trend than the inner part. This signifies that the outer part of the galaxy mostly has younger sources, while the inner part shows a uniform distribution.

\subsection{Age estimation}
\begin{figure}
    \centering
    \includegraphics[width=3.7in]{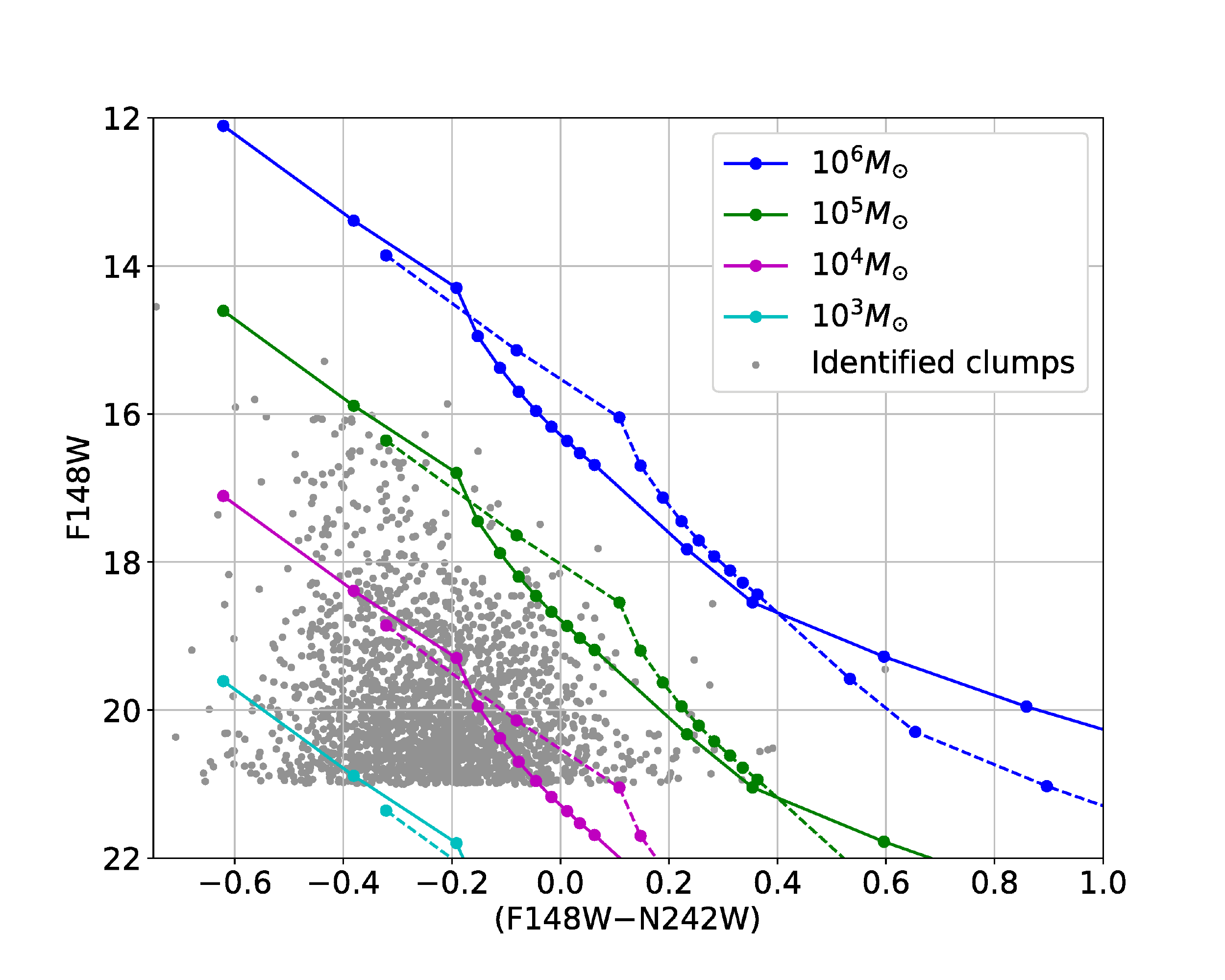}
    \caption{The identified star-forming clumps are over plotted (grey filled circles) on the simulated model curves as shown in Figure \ref{starburst99}.}
    \label{cmd_galaxy}
\end{figure}

\begin{figure}
    \centering
    \includegraphics[width=3.7in]{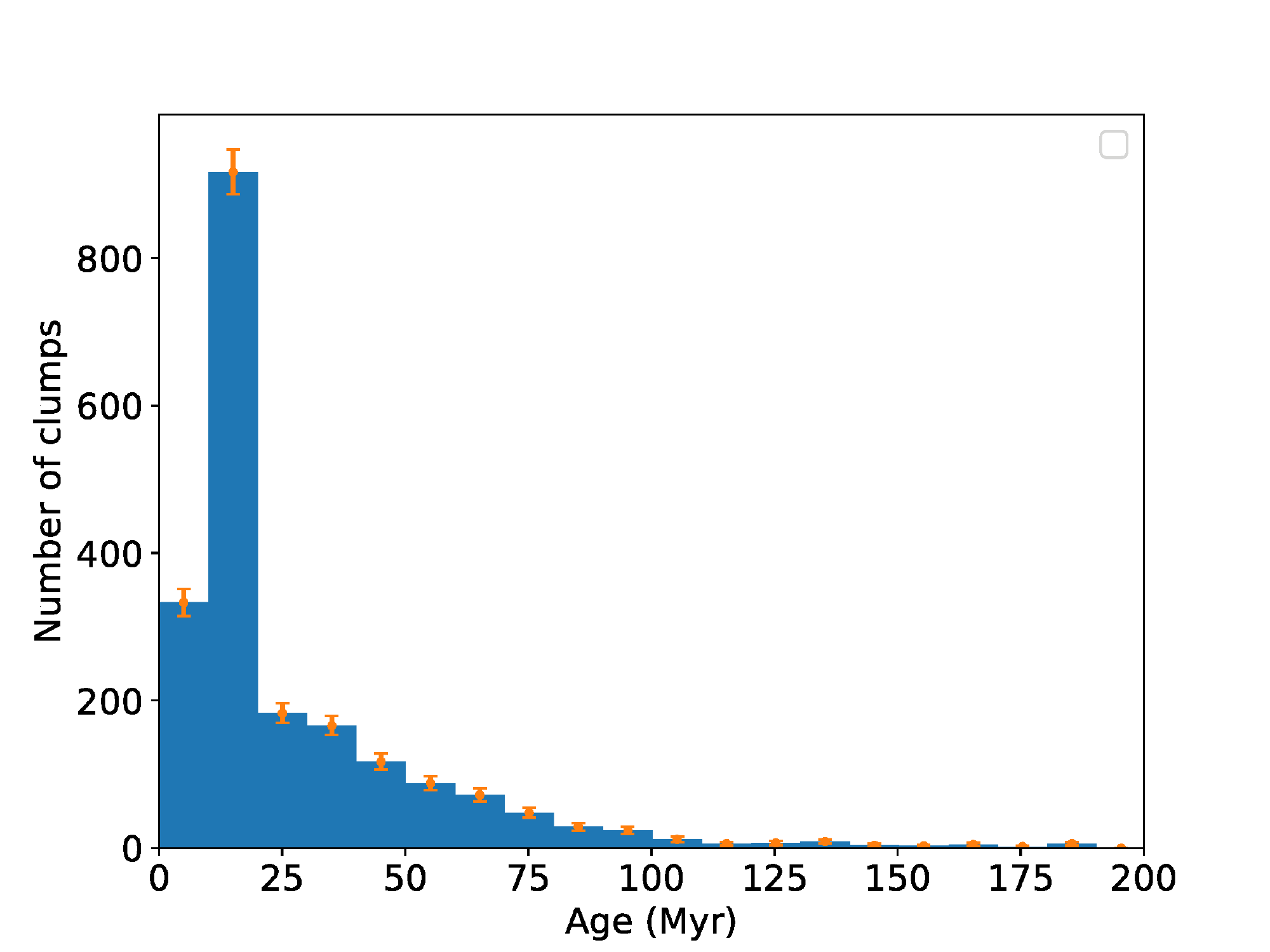}
    \caption{The age histogram of the identified star-forming clumps.}
    \label{age_hist}
\end{figure}

As discussed in Section \ref{s_model}, the age of an unresolved star cluster can be estimated from its observed (F148W$-$N242W) color. The child structures identified in the FUV image are likely to be the unresolved FUV bright star-forming clumps of the galaxy. Among all, the smaller clumps are mostly composed of a single entity, whereas the larger ones can be stellar associations or a combination of multiple clumps, which could not be resolved further with UVIT. In order to apply the SSP model to these clumps, we assumed each of them as single entities. We estimated the background and extinction corrected F148W magnitude and (F148W$-$N242W) color of these identified clumps and showed in Figure \ref{cmd_galaxy} (grey points) along with the model curves of Figure \ref{starburst99}. Only a few observed points are lying outside the model color range on the bluer side. This may arise due to the overestimation of extinction or large photometric error. We have excluded these points from our analysis. We considered (F148W$-$N242W) color for rest of the clumps and performed linear interpolation along the color axis to estimate age of each clump. The histogram of the estimated ages is shown in Figure \ref{age_hist}. We noticed a richness of younger clumps with age less than 20 Myr. This suggests that the stellar populations identified in the FUV image are mostly younger, and the galaxy has undergone an enhanced phase of star formation in the last 20 Myr.

\begin{figure*}
    \centering
    \includegraphics[width=7in]{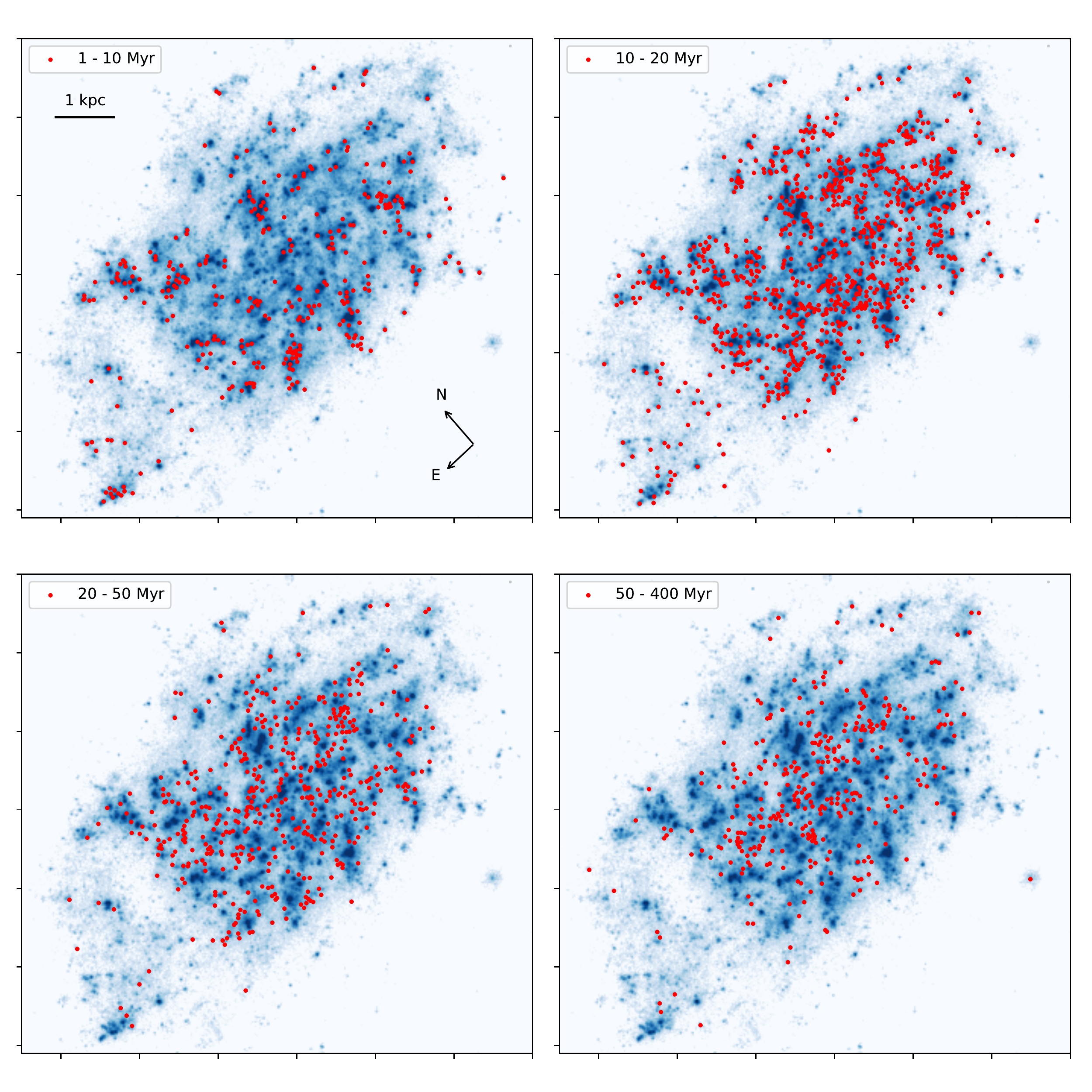}
    \caption{Spatial distribution of clumps (red points) as a function of their estimated age between the range 1 - 400 Myr. The figures represent four different age groups as mentioned in the text. The age range of each group is mentioned in the corresponding panel. The background image shows the F148W band flux map of the galaxy.}
    \label{age_dist1}
\end{figure*}

\subsection{Age distribution}
\label{age_dist_ngc7793_s}
The spatial distribution of star-forming clumps as a function of age conveys the star formation history across a galaxy. As star formation in a galaxy can be triggered by multiple mechanisms, the spatial age map of the young star-forming clumps is important to draw conclusions about the impact of the local environment and also the possible triggering activities. This will also shed light on the nature of disk growth in the galaxy. To visualize the spatial distribution of the identified star-forming clumps as a function of age in NGC~7793, we divided them into four different groups between the age range 1 - 400 Myr. The age range of the groups are 1 - 10 Myr, 10 - 20 Myr, 20 - 50 Myr, and 50 - 400 Myr. The bin size of the groups is chosen on the basis of age histogram shown in Figure \ref{age_hist}. We fixed the interval to be smaller where there is more number of clumps and to be wider where there is less number of clumps. Also, for each selected bin, the value of the mean error in the estimated ages is less compared to the bin size. 

The locations of the clumps identified in each age group are shown on the galaxy image in Figure \ref{age_dist1}. The youngest star-forming clumps (age group: 1 - 10 Myr) are mostly found to be located along the flocculent arms in the outer part of the galaxy. The distribution of these clumps is also more compact in nature. In the case of clumps with age between 10 -20 Myr, we noticed a similar pattern with a predominant distribution along the arms. We find an overall scattered distribution of clumps in the age range 20 -50 Myr across the galaxy. These are preferably found away from the arms. Beyond the age 50 Myr, the clumps are noticed to be present more in the inner part of the galaxy. The regions between two arms are mostly populated by clumps older than 20 Myr. The overall picture thus suggests that star formation in the last 20 Myr is taking place mostly along the flocculent arms of the galaxy. The star-forming regions located in the far outer part in the east direction are found to be populated with clumps younger than 20 Myr.

\subsection{Mass estimation}
\begin{figure}
    \centering
    \includegraphics[width=3.5in]{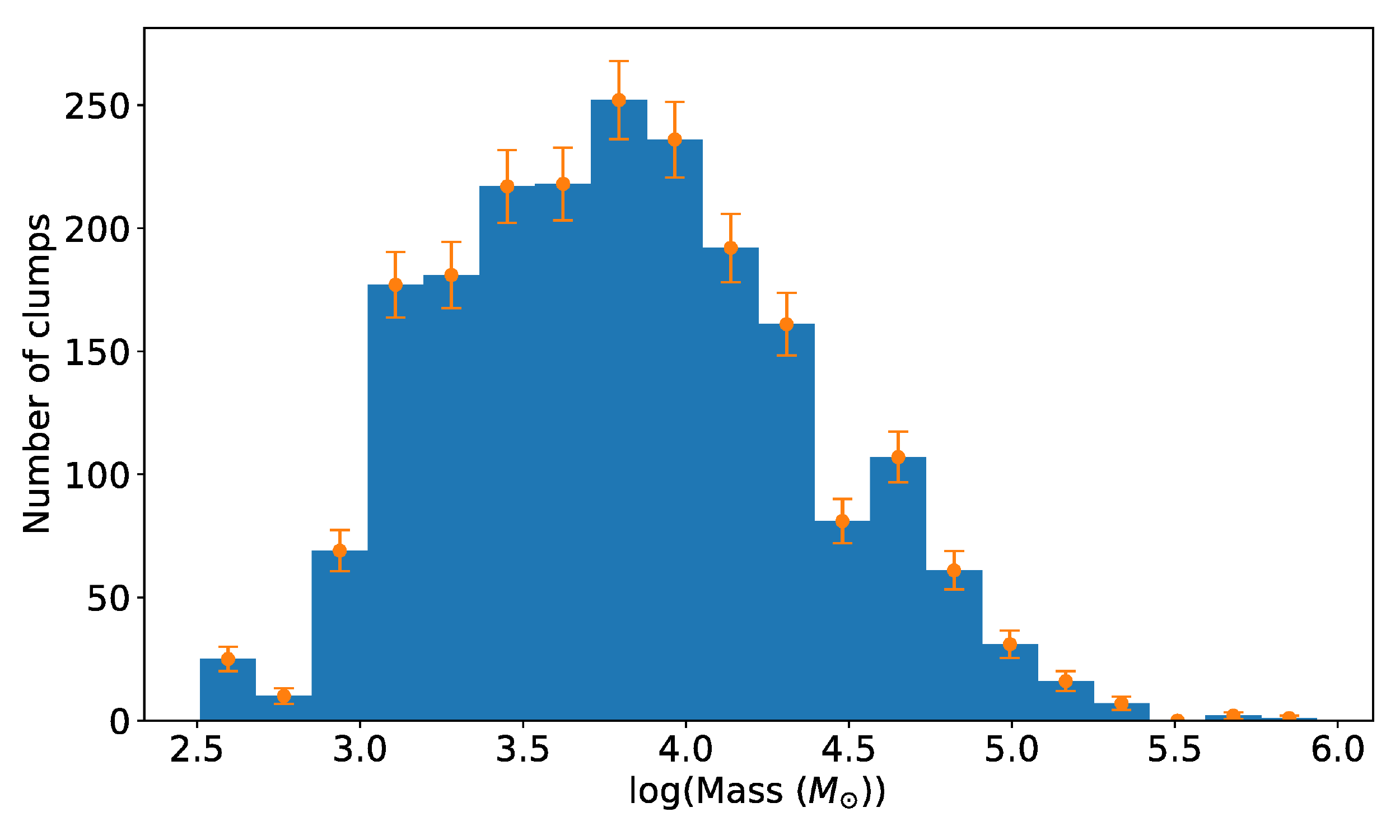}
    \caption{The mass histogram of the identified star-forming clumps.}
    \label{mass_hist}
\end{figure}

We also estimated mass of each clump from its observed F148W magnitude. For the known value of (F148W$-$N242W) color of each clump, we used the F148W band observed magnitude and performed a linear interpolation along the magnitude axis of Figure \ref{cmd_galaxy} to estimate mass. The histogram for the estimated mass is shown in Figure \ref{mass_hist}. It shows that the identified clumps have a mass range between $3\times10^2 M_{\odot}$ - $10^6 M_{\odot}$ with a peak around $10^4 M_{\odot}$. The GMCs identified in this galaxy also has a similar mass range \citep{grasha2018}. The estimation of masses below $10^3 M_{\odot}$ is not accurate due to the lower limit of the model mass range. The majority of the clumps are found to have an intermediate-mass between $10^3 M_{\odot}$ - $10^4 M_{\odot}$. The number of clumps with mass greater than $10^5 M_{\odot}$ is relatively small in number. This might indicate that the galaxy has not formed much of massive complexes in recent times.

\subsection{Mass distribution}
\begin{figure*}
    \centering
    \includegraphics[width=7in]{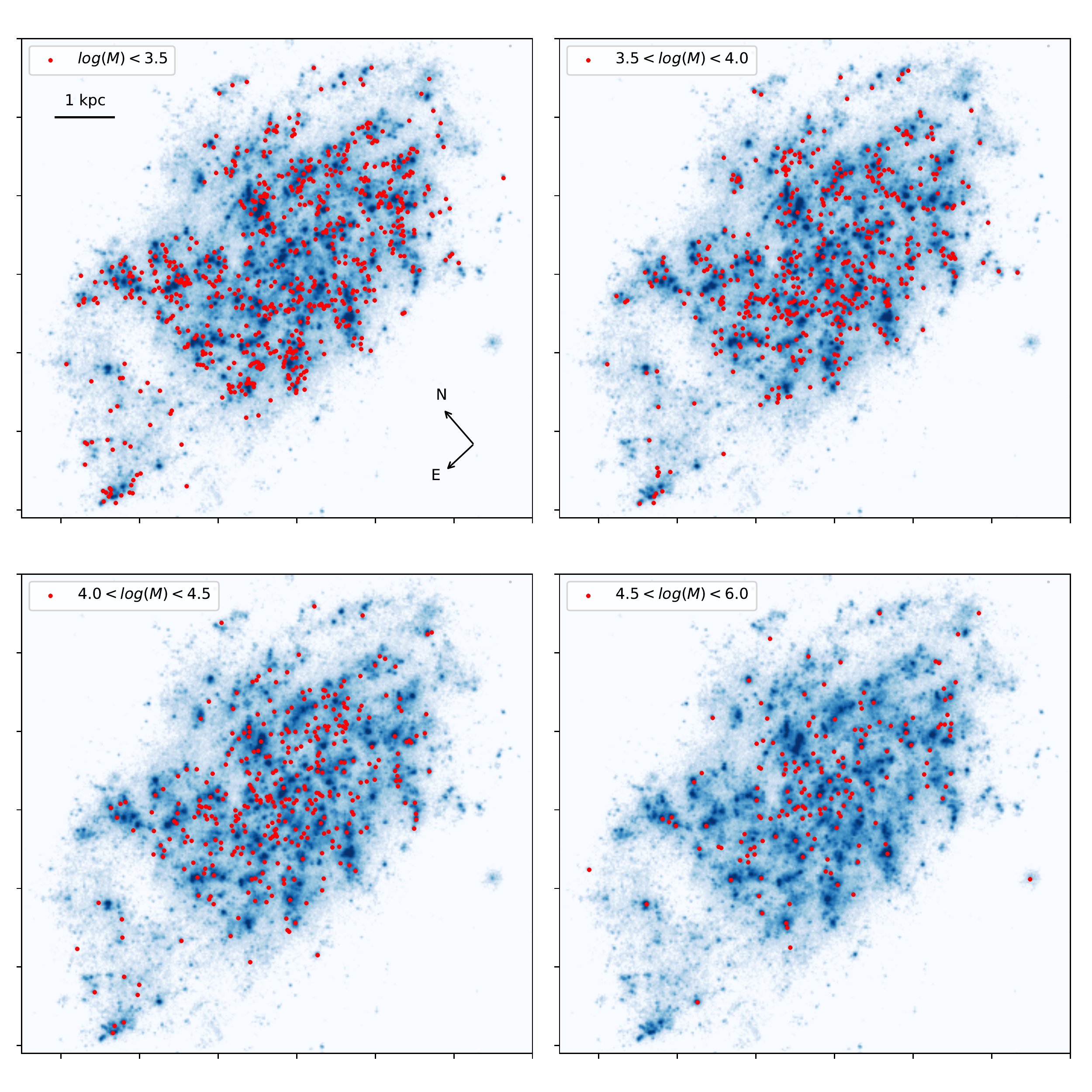}
    \caption{Spatial distribution of clumps (red points) as a function of their estimated mass between the range $3\times10^2 - 10^6$ M$_{\odot}$. The figures represent four different mass groups as mentioned in the text. The mass range of each group is mentioned in the corresponding panel. The background image shows the F148W band flux map of the galaxy.}
    \label{mass_dist1}
\end{figure*}

The mass of a star-forming clump depends on the mass of the parent molecular cloud from where it has formed. The massive clumps generally form from giant molecular clouds, whereas low mass clumps can be produced from molecular cloud of relatively smaller mass. It is thus important to know the mass distribution of clumps to explore the star-forming environment across a galaxy. In order to understand the mass distribution of identified clumps across the galaxy, we made four groups for different mass ranges. The mass ranges of the groups are log(M/M$_{\odot}$) $<$ 3.5, 3.5 $<$ log(M/M$_{\odot}$) $<$ 4.0, 4.0 $<$ log(M/M$_{\odot}$) $<$ 4.5 and 4.5 $<$ log(M/M$_{\odot}$) $<$ 6.0. The bins are fixed by following the same steps, as mentioned in Section \ref{age_dist_ngc7793_s}. In Figure \ref{mass_dist1}, we have shown the position of these clumps on the FUV image of the galaxy. We noticed a hierarchical distribution of star-forming clumps as a function of mass. The low mass clumps (log(M/M$_{\odot}$) $<$ 3.5) are mostly distributed along the flocculent arms. The central region of the galaxy has a few of these low mass clumps. The clumps, with mass in the range 3.5 $<$ log(M/M$_{\odot}$) $<$ 4.0, show similar distribution with relatively less number. These clumps are seen more in the inner part of the arms than its outer part. The distribution of more massive clumps with mass between 4.0 $<$ log(M/M$_{\odot}$) $<$ 4.5 gradually shrinks towards the inner part. These are also seen in between the arms. The most massive clumps (4.5 $<$ log(M/M$_{\odot}$) $<$ 6.0) of the galaxy are distributed more in the central part of the galaxy. The overall scenario portrayed that the inner part of the flocculent arms contains both low and high mass clumps, whereas the outer part of these arms is populated with more of low mass clumps with mass log(M/M$_{\odot}$) $<$ 3.5. This brings out the hierarchical distribution of star-forming clumps along the spiral arms.

\subsection{Nuclear star cluster}
Nuclear star cluster is known to be a dense stellar system mostly seen in the dynamical center of disk galaxies \citep{neumayer2020,seth2019}. The sizes of these objects are similar to those of globular clusters. It has been found that more than 75\% of nearby late-type spiral galaxies have a nuclear star cluster at their center \citep{boker2002}. NGC~7793 is also reported to have a nuclear star cluster \citep{walcher2006,carson2015,kacharov2018}. The cluster has an effective radius of 12.45 pc, measured in the HST F275W band \citep{carson2015}. \citet{carson2015} studied the cluster with multi-band HST data and reported that the size of the cluster is bigger in U band than in optical. The SFH of the cluster shows a complex nature. The Very Large Telescope (VLT) spectroscopic observations by \citet{kacharov2018} found that the cluster contains stellar population of four different age ranges. The majority of the cluster populations are older than 10 Gyr, whereas some have age around 2 Gyr, some between 200 - 600 Myr, and some are very young with age $\sim$ 10 Myr. 

\begin{figure}
    \centering
\includegraphics[width=3.5in]{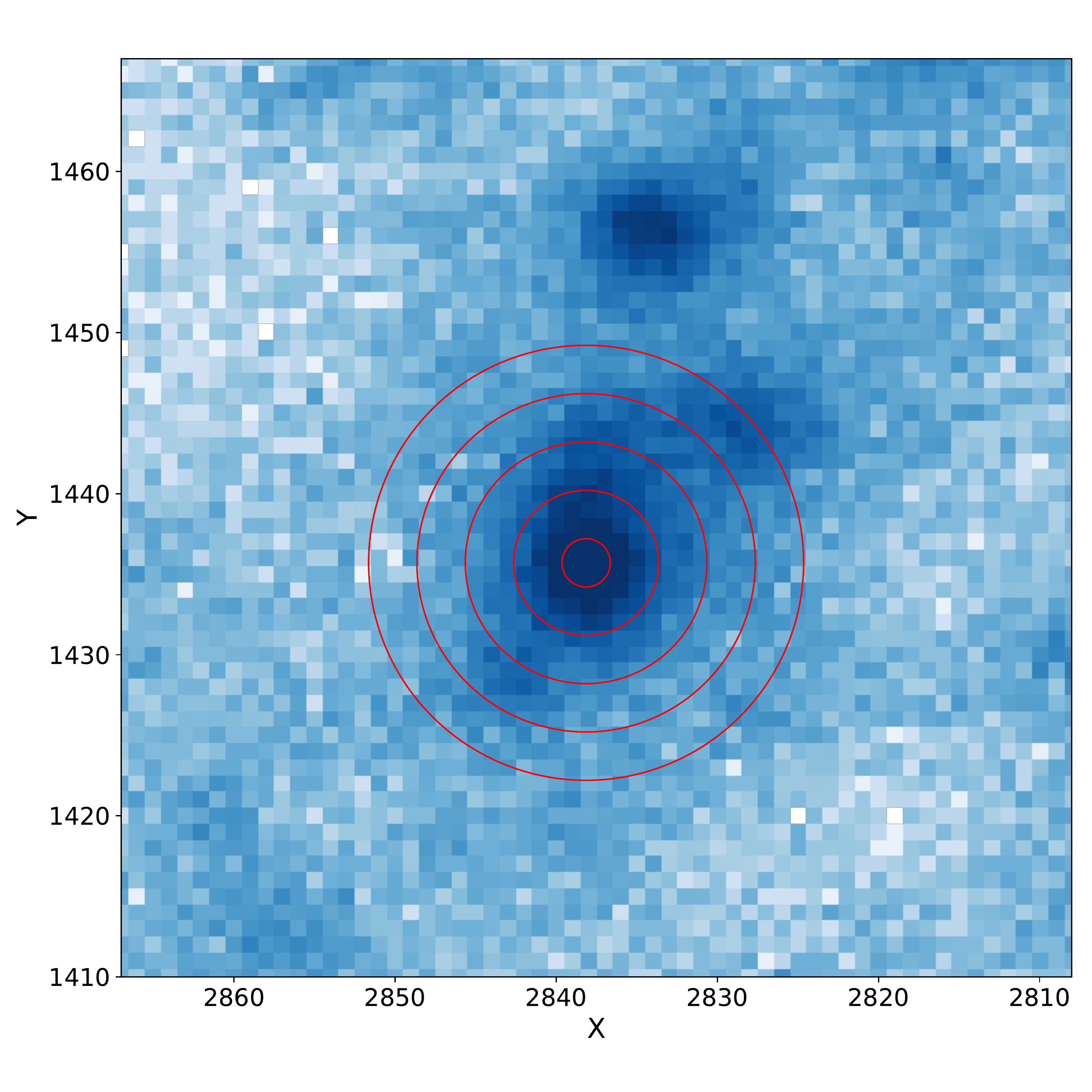}
\includegraphics[width=3.7in]{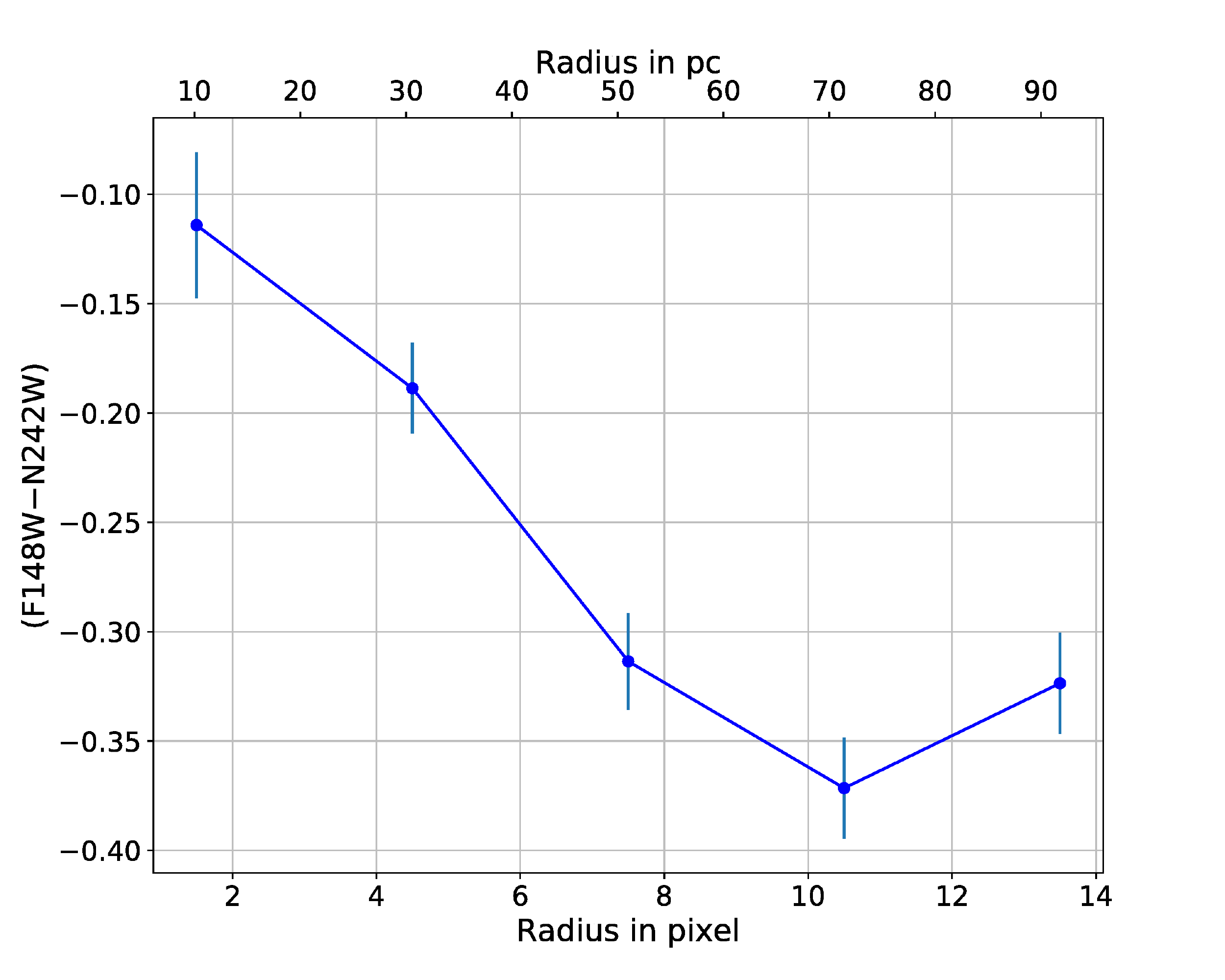}
    \caption{The upper panel (a) shows the nuclear star cluster as seen in the UVIT FUV image. The red circles represent five different apertures with radius starting from 1.5 pixel to 13.5 pixel. The lower panel (b) shows the extinction and background corrected (F148W$-$N242W) color profile of the object as measured from the annuli displayed above.}
    \label{nsc}
\end{figure}

In this study, we used UVIT FUV and NUV observations to characterize the nuclear cluster. To check the properties in UV, we considered both FUV and NUV images and defined multiple apertures centered on the reported cluster position. The radius of the smallest aperture is considered as 1.5 pixel ($\sim$ 0.6$^{\prime\prime}$, i.e., diameter of the aperture $\sim$ FWHM of PSF) and further increased with an increment of 3 pixel to define four more apertures (Figure \ref{nsc}a). We measured fluxes in both F148W and N242W bands for each of these defined annuli. The measured fluxes are corrected for extinction and background to estimate (F148W$-$N242W) color which is shown in Figure \ref{nsc}b as a function of aperture radius. The (F148W$-$N242W) color gradually becomes bluer with increasing radius and flattens beyond a radius of 10 pixels. This signifies that the stellar populations present in the outskirts of the nuclear cluster are younger than the inner part, which indirectly supports the earlier conclusion of circum-nuclear star formation by \citet{carson2015}. This can also happen due to the accretion of the younger population from the nearby stellar groups to the nuclear cluster. The variation in (F148W$-$N242W) color in the nuclear region can be unreal if there exists a significant variation in extinction. The study by \citet{kahre2018} showed that extinction in the central region of the galaxy is relatively high, and it has a nearly constant value. This rules out the possibility of color variation due to the variation in extinction. 

We also identified this cluster as a single star-forming clump with the method as discussed in section \ref{clumps_7793_s}. The estimated values of cluster age and mass are found to be 19.1$\pm$0.8 Myr and 2.3$\times10^5 M_{\odot}$ respectively. As FUV emission traces the younger populations, this age and mass values, therefore, sample the properties of the stellar populations, that are the youngest among the four different types as found by \citet{kacharov2018}.

\section{Results and Discussion}
\label{s_results}
The primary aim of this study is to understand the UV disk structure of NGC~7793 and to identify young star-forming clumps in the galaxy and estimate their age and mass. To do that, we used FUV and NUV imaging data of the galaxy, observed with F148W and N242W filters of UVIT. The spatial resolution of UVIT has helped to identify clumps of radii up to $\sim$ 12 pc. The fluxes measured in FUV and NUV bands are compared with the model values to estimate ages and masses of the clumps.

We used Starburst99 model data to simulate the diagnostic plot which helped us estimating age and mass of the clumps from their observed F148W magnitude and (F148W$-$N242W) color. Therefore, errors in the measurements of magnitude and color will reflect in the estimated values of age and mass. The photometric error in both F148W and N242W bands lie within a range between $\sim$ 0.01 - 0.1 mag. The corresponding error in age has a range between $\sim$ 1 - 80 Myr. The younger clumps have less error in age, while the older clumps have relatively large errors. The mean error value for each of the four age groups defined in our study have value smaller than the bin size. For example, the group containing clumps with age between 1 - 10 Myr has a mean error of $\sim$ 3 Myr. Similarly, the error in F148W magnitudes results in $\sim$ 1 - 15\% error in the estimated mass values. Another parameter that can change the measured age and mass values of the clumps is the adopted value of E(B$-$V). The extinction and reddening inside a galaxy can have spatial variation, which will affect the observed flux differently in different locations. Therefore, our assumption of a fixed reddening value across the galaxy have added some amount of uncertainty in the estimated values of age and mass. A larger reddening will make the clumps younger and massive and vice versa. We adopted the extinction law to be an average LMC type for dealing with the interstellar extinction in NGC~7793. A change in extinction law will also affect the estimated age and mass values accordingly. The identified star-forming regions of the galaxy may also have different metallicity, and this can further affect the derived parameters as color changes with varying metallicity for a fixed age.

Several studies have been done with observations in different bands to understand the disk properties of NGC~7793. \citet{vlajic2011} observed two fields in the extended outer part of the galaxy with Gemini Multi Object Spectrograph (GMOS) and reported that the galaxy disk beyond $\sim$ 5 kpc is mostly populated with older red giant branch (RGB) stars. Their study could trace the stellar disk up to $\sim$ 10 kpc. With HST observations, \citet{radburn2012} studied a part of the galaxy disk in the eastern side and found that the surface density of the older population in the outer disk between 3 - 5 kpc decreases gradually while it remains almost constant in the case of younger population. They also noticed a break in the distribution of younger stars at a radius of 5 kpc. The luminosity density profiles presented in our study support the earlier results. The FUV luminosity profile, derived from UVIT observation, also shows a break at radius 5 kpc. We further noticed the FUV profile to take over NUV at 3 kpc radius and becomes more flat up to 5 kpc radius. We also found that the clumps identified between radii 3 -5 kpc are mostly bluer. This signifies that the stellar populations identified in UV in the outer disk between 3 - 5 kpc are mostly younger. It is also possible that the outer disk of the galaxy has relatively low reddening than the adopted value. A decrease in the E(B$-$V) value will make the clumps redder, which will make the color distribution of clumps in the outer disk, between radii 3 - 5 kpc, flatter. 

The drop in the FUV luminosity beyond 5 kpc conveys that the disk outside this radius has less number of younger and/or massive population. Both the work by \citet{vlajic2011} and \citet{radburn2012} studied a limited part of the disk, whereas our study covered the entire galaxy disk from center to outside. The radial nature of the disk, derived from a limited region in the eastern side by HST, found to remain similar when averaged over the entire disk. Hence the enhancement in the recent star formation between radius 3 - 5 kpc has happened mostly along all azimuthal directions. \citet{sacchi2019} performed a more comprehensive study by almost covering the entire disk with HST observations to understand the radial star formation history in the galaxy. They studied the distribution of stars of age from a few Myr to $\sim$ 10 Gyr and reported an inside-out growth for the galaxy disk. The SFR in the outer part of the galaxy has been enhanced in more recent times while the inner part is mostly populated with populations older than 1 Gyr. We measured the values of disk scale-length in FUV and NUV as 2.64$\pm$0.16 kpc and 2.21$\pm$0.21 kpc respectively. The optical scale-length of the galaxy is reported as 1.08 kpc by \citet{carignan1985}. These together signify that the disk becomes extended in shorter wavelengths. This, in other words, means that the older populations are more centrally concentrated, which portrays an inside-out growth scenario for the disk of NGC~7793. The HST observations presented by \citet{grasha2018} and \citet{sacchi2019} have not covered the outskirts of the galaxy disk. We detected star-forming clumps younger than 20 Myr in such regions which signifies recent star-forming activity in the disk outskirts. The larger star-forming knot located in the eastern outskirt (which was not covered by HST) can be seen to undergo recent star formation. This further strengthens the proposition of inside-out disk formation of the galaxy.

The H~I disk of the galaxy also shows some noticeable characteristics. \citet{carignan1990} noticed non-circular motion in the northern part of the galaxy and speculated the possibility of past interactions with a nearby companion of the sculptor group. Along the north and north-eastern part, we noticed some parts of the star-forming UV disk to extend outside the extent of H~I gas of density more than $10^{21} cm^{-2}$. These features could be a result of a recent interaction that resulted in the enhancement of star formation in the outer part of the galaxy.

There are ample evidence of recent star-forming activity in NGC~7793. An HST study done by \citet{grasha2018} covered almost the entire galaxy disk up to a radius of around 5 kpc and identified 293 young star clusters, with 65\% of them to be younger than 10 Myr. \citet{bibby2010} identified 74 emission-line regions across the galaxy disk using imaging observation from VLT/FOcal Reducer and Spectrograph (FORS1). \citet{dicaire2008} detected H$\alpha$ emission up to the edge of the H~I gas disk of the galaxy and speculated massive stars as the source of this emission. These all together convey an enhancement in the recent star formation across the galaxy. In our study, we found around 61\% of the 2046 FUV identified clumps to be younger than 20 Myr, which signifies enhanced star-forming activity in the galaxy during the last 20 Myr. 

\citet{elmegreen2014} studied the galaxy as a part of the HST Legacy UV survey and reported a hierarchical distribution of star-forming regions with a range of size between $\sim$ 1 - 70 pc. The star-forming clumps identified in our study found to have radii in the range 12 - 70 pc. The clumps with radii smaller than 12 pc could not be resolved in our study due to the resolution limit of UVIT. It is also possible that the larger clumps identified by UVIT are actually a combination of multiple smaller clumps, which appear as a single clump in the UVIT images. \citet{grasha2018} reported a similar range for the radius of GMCs identified in the galaxy. They further concluded that the younger clusters are more closely associated with GMCs while the older ones are found to be more dispersed away from the natal cloud. This, along with the similarity in sizes, further strengthens the connection between star-forming clumps and molecular clouds in the galaxy. We also searched for the largest star-forming parent structure identified by {\it astrodendro} for the adopted threshold flux and found that it has a size of $\sim$ 3 kpc. This length scale, which is sensitive to the value of threshold flux, will become smaller with increasing threshold and vice versa.

Our study found that the distribution of youngest star-forming clumps (Age $<$ 10 Myr) is more compact in nature, and they trace the flocculent arms of the galaxy. This signifies that the enhancement of star formation during the last 10 Myr has specifically happened along the arms. A recent study by \citet{sacchi2019} found that the presence of a spiral density wave is not clearly seen in the distribution of stellar populations older than 1 Gyr in the galaxy, whereas the younger populations were mostly seen along the flocculent arms. Following this picture, they inferred about the possible lack of spiral density waves in the galaxy. In this study, using UV data we traced only the young star-forming clumps up to age $\sim$ 400 Myr and noticed the youngest stellar clumps to distinctly trace the flocculent arms of the galaxy. The role of density waves in triggering star formation in disk galaxies is still among the debated topics. Therefore, these results offer a scenario to explore whether the arm structures observed in our study are only due to the distribution of star-forming regions or these are formed due to the impact of spiral density waves which has triggered star formation along the flocculent arms.

Our study also reveals that the central region of the galaxy does not show much star formation during the last 10 Myr. This justifies well with the fact that the central 1 kpc region of the galaxy has much lower molecular gas concentration (n$_{H_2}\sim$ 10$^{2.1}$ cm$^{-3}$) than the global average as reported by \citet{muraoka2016}. On the other hand, the molecular gas density is found to be relatively higher away from the center, specifically along two arms in the east and west directions. We identified several star-forming clumps younger than 10 Myr in these locations, which points to the active star formation in regions with dense molecular gas.

The star-forming clumps identified in our study mostly cover a mass range between $10^3 - 10^5 M_{\odot}$. We noticed only a few clumps more massive than $10^5 M_{\odot}$. This picture matches with the mass of GMCs identified in the galaxy \citep{grasha2018}. The clumps with relatively more mass (4 $<$ log(M/M$_{\odot}$) $<$ 6) are mostly seen in the inner part of the galaxy disk. This can be the result of an artifact due to crowding. The inner part of a disk galaxy has a more crowded environment than the outer part, and therefore, it is possible that some of the massive clumps identified in the inner disk are actually a group of multiple clumps that could not be resolved by UVIT. The other interesting result is the distribution of clumps along the flocculent arms. We noticed that the ends of the arms are populated with more low mass clumps (log(M/M$_{\odot}$) $<$ 3.5), whereas the inner part of arms, have low as well as more massive clumps. This portrays a gradient in the mass distribution from inner to outer part along the arms. 

We have also characterized the nuclear star cluster of the galaxy with UV observations. \citet{carson2015} reported that the nuclear cluster of NGC~7793 has a decreasing effective radius with increasing wavelength, which signifies a recent circum-nuclear star formation. \citet{walcher2006} found stellar population younger than 100 Myr in the nucleus of the galaxy. \citet{kacharov2018} reported a complex star formation history for this cluster. They noticed stellar population of different ages starting from $\sim$ 10 Myr to $>$ 10 Gyr as part of the cluster. They concluded that merging of multiple clusters with different ages could give rise to the observed properties of the nuclear cluster. The (F148W$-$N242W) color profile of the cluster, derived in our study, found to become bluer with increasing aperture size. This supports the possibility of circum-nuclear star formation or accretion of young stellar population from nearby stellar groups to the nuclear cluster. The effective radius of the cluster is reported to be 12.45 pc in the HST F275W band \citet{carson2015}. Considering the limit of UVIT resolution, the smallest aperture we could define is around this value. Therefore, the apertures larger than the effective radius (i.e., the annuli in Figure \ref{nsc}) of the cluster basically trace stellar population in the cluster outskirts or those which are in the process to be accreted by the nuclear cluster. We estimated the age and mass of the cluster as 19.1$\pm$0.8 Myr and 2.3$\times10^5 M_{\odot}$ respectively. As FUV and NUV emission primarily trace stellar population of age up to a few hundred Myr, the estimated age of the cluster, which reported to have stellar populations of a wide age range, characterizes the younger populations. Similarly, the estimated mass of the cluster signifies the amount of mass contributed by younger populations present in the cluster. 

\section{Summary}
\label{s_summary}
The main results of this study are summarised below.
\begin{enumerate}
\item We used UVIT FUV and NUV observations to understand the UV disk emission profile and to identify young star-forming clumps in the galaxy NGC~7793.
\item The value of the galaxy disk scale-length ($R_{d}$) is found to increase towards shorter wavelengths from optical to FUV. We estimated the value of $R_{d}$ to be 2.64$\pm$0.16 kpc and 2.21$\pm$0.21 kpc in FUV and NUV respectively. 
\item Relative to the inner part, we noticed FUV emission to be more dominant than the NUV emission in the outer part of the galaxy (between radius 3 - 5 kpc).
\item We identified 2046 young star-forming clumps in the galaxy with radii ranging between $\sim$ 12 - 70 pc.
\item The majority of the clumps were found to have age younger than 20 Myr, which signifies enhancement in recent star formation across the galaxy.
\item The youngest clumps (age $<$ 10 Myr) are specifically found along the flocculent arms of the galaxy. 
\item The identified clumps cover a mass range between $10^3 M_{\odot} - 10^6 M_{\odot}$.
\item We noticed a hierarchical mass distribution of the clumps along the flocculent arms. The end of the flocculent arms has more low mass clumps, whereas the inner part of those arms contains low as well as more massive clumps.
\item The extent of the star-forming UV disk of the galaxy closely matches with the extent of H~I disk with column density more than $10^{21} cm^{-2}$. 
\item The stellar population in the outskirts of the nuclear star cluster is found to be younger compared to the inner part. This signifies a possibility of circum-nuclear star formation or accretion of the younger stellar population from nearby stellar groups to the nuclear cluster.
\end{enumerate}

\acknowledgments
UVIT project is a result of collaboration between IIA, Bengaluru, IUCAA, Pune, TIFR, Mumbai, several centres of ISRO, and CSA. Indian Institutions and the Canadian Space Agency have contributed to the work presented in this paper. Several groups from ISAC (ISRO), Bengaluru, and IISU (ISRO), Trivandrum have contributed to the design, fabrication, and testing of the payload. The Mission Group (ISAC) and ISTRAC (ISAC) continue to provide support in making observations with, and reception and initial processing of the data.  We gratefully thank all the individuals involved in the various teams for providing their support to the project from the early stages of the design to launch and observations with it in the orbit. This research made use of Matplotlib \citep{matplotlib2007}, Astropy \citep{astropy2013,astropy2018}, Astrodendro (http://www.dendrograms.org/), community-developed core Python packages for Astronomy. Finally, we thank the referee for valuable suggestions.

\software{CCDLAB \citep{postma2017}, SAOImageDS9 \citep{joye2003}, Matplotlib \citep{matplotlib2007}, Astropy \citep{astropy2013,astropy2018}, Astrodendro (http://www.dendrograms.org/)}

%\bibliographystyle{aasjournal}
%\bibliography{reference}

\end{document}